\documentclass{aastex62}

\received{ }
\revised{ }
\accepted{ }
\submitjournal{ApJS}

\usepackage{xcolor}

\shorttitle{{\bf Hot subdwarfs in Gaia DR2 and LAMOST DR7}}
\shortauthors{Luo et al.}

\begin{document}

\title{Hot subdwarf Atmospheric parameters, Kinematics, and Origins -- Based on 1587 hot subdwarf stars observed in Gaia DR2 and LAMOST DR7}

\correspondingauthor{Yangping Luo}
\email{ypluo@bao.ac.cn}

\author{Yangping Luo}
\affiliation{Department of Astronomy, China West Normal University, \\
Nanchong, 637002, PR China}
\author{P\'{e}ter~N\'{e}meth}
\affiliation{Astronomical Institute of the Czech Academy of Sciences, Fri\v{c}ova 298, CZ-251\,65 Ond\v{r}ejov, Czech Republic}
\affiliation{Astroserver.org, F\H{o} t\'er 1, 8533 Malomsok, Hungary}
\author{Kun Wang}
\affiliation{Department of Astronomy, China West Normal University, \\
Nanchong, 637002, PR China}
\author{Xi Wang}
\affiliation{Department of Astronomy, China West Normal University, \\
Nanchong, 637002, PR China}
\author{Zhanwen Han}
\affiliation{Key Laboratory for the Structure and Evolution of Celestial Objects, Chinese Academy of Sciences, \\
Kunming 650011, PR China}

\begin{abstract}
Based on the Gaia DR2 catalogue of hot subdwarf star candidates, we identified 1587 hot subdwarf stars with spectra in LAMOST DR7.
We present atmospheric parameters for these stars by fitting the LAMOST spectra with {\sc Tlusty/Synspec} non-LTE synthetic spectra. Combining LAMOST radial velocities and Gaia Early Data Release 3 (EDR3) parallaxes and proper motions, we also present the Galactic space positions, velocity vectors, orbital parameters and the Galactic population memberships of the stars.
With our He classification scheme, we identify four groups of He rich hot subdwarf stars in the $T_{\rm eff}-\log\,g$ and $T_{\rm eff}-\log{(n{\rm He}/n{\rm H})}$ diagrams.
We find two extreme He-rich groups ($e$He-1 and $e$He-2) for stars with $\log{(n{\rm He}/n{\rm H})}\geq0$ and two intermediate He-rich groups ($i$He-1 and $i$He-2) for stars with $-1\le\log{(n{\rm He}/n{\rm H})}<0$.
We also find that over half of the stars in Group $e$He-1 are thick disk stars, while over half of the stars in Group $e$He-2 correspond to thin disk stars.
The disk population fractions of Group $i$He-1 are between those of Group $e$He-1 and $e$He-2. Almost all stars in Group $i$He-2 belong to the thin disk.
These differences indicate that the four groups probably have very different origins. Comparisons between hot subdwarf stars in the halo and in the Galactic globular cluster $\omega$ Cen show that only He-deficient stars with $-2.2\le\log{(n{\rm He}/n{\rm H})}<-1$ have similar fractions.
Hot subdwarfs with $\log{(n{\rm He}/n{\rm H})}\ge 0$ in $\omega$ Cen have no counterparts in the thick disk and halo populations, but they appear in the thin disk.
\end{abstract}

\section{Introduction} \label{sec:intro}
Hot subdwarf stars are evolved core He burning stars with thin H envelopes on the extreme horizontal branch (EHB,  \citealt{2009ARA&A..47..211H,2016PASP..128h2001H}).
Their spectral features are similar to that of typical O/B type stars, but their luminosities are orders of magnitudes lower.
They are classified according to their spectral features into sdB and sdO types \citep{2013A&A...551A..31D}.
Hot subdwarf stars are interesting objects in many fields of astronomy.
They are found in all galactic populations, and they are known to be the main source of the ultraviolet excess radiation of elliptical galaxies and the bulges of spiral galaxies \citep{2007MNRAS.380.1098H}, and also responsible for the extended horizontal branch morphology of globular clusters \citep{2008A&A...484L..31H, 2013A&A...549A.145L, 2015MNRAS.449.2741L}.
Hypervelocity hot subdwarf stars allow us to probe the Galactic gravitational potential and put constraints on the mass of the Galactic dark matter halo \citep{2015Sci...347.1126G, 2016ApJ...821L..13N}.
The orbital-period - mass-ratio (P-q) relation in wide hot subdwarf binaries reflects the Galactic history \citep{2020A&A...641A.163V}.
Hot subdwarf binaries with massive compact companions play a role as progenitors of cosmologically important type Ia supernovae \citep{2009A&A...493.1081J, 2009MNRAS.395..847W, 2010A&A...515A..88W, 2013A&A...554A..54G, 2015A&A...577A..26G,  2020ApJ...902...92R} and mark potential gravitational-wave sources, that may be strong enough to be detectable by future space-based missions, such as the Laser Interferometer Space Antenna \citep{2018A&A...618A..14W, 2020A&A...634A.126W}.

Hot subdwarf stars are excellent probes for binary evolution, stellar atmosphere and interior models.
Because a large fraction of sdB stars must have gone through a common-envelope (CE) phase of evolution and are detached binaries, they provide a clean-cut laboratory to study this poorly understood, yet crucial phase of binary evolution and tidal effects \citep{2013A&ARv..21...59I, 2020A&A...642A.180P,2020RAA....20..161H}.
Hot subdwarf stars display strong chemical peculiarities, in particular for the He abundance, which may be a trace element as well as the dominant constituent of the atmosphere, depending on the hot subdwarf type \citep{2003A&A...400..939E,2012MNRAS.427.2180N,2018MNRAS.475.4728B}.
Furthermore, the atmospheres of a few intermediate He-rich hot subdwarfs show extreme heavy metal abundances, with abundances of lead reaching nearly 10\,000 times the solar value, that is 10-100 times higher than that previously measured in normal hot subdwarf atmospheres \citep{2011MNRAS.412..363N, 2013MNRAS.434.1920N, 2020MNRAS.491..874N, 2017MNRAS.465.3101J, 2019MNRAS.489.1481J, 2018A&A...609A..89L, 2019MNRAS.482..758S, 2019A&A...630A.130D, 2020A&A...643A..22D, 2020MNRAS.496.2558F}.
Several types of pulsating stars have been discovered among hot subdwarfs, being remarkable targets for asteroseismology to probe their interior structure \citep{2010A&A...516L...6C, 2018ApJ...853...98Z, 2019ApJ...878L..35K, 2020MNRAS.495.2844S, 2020MNRAS.496..718J, 2014A&A...569A..15O, 2020MNRAS.499.3738O}.

Despite their importance in many fields of astronomy, the formation of hot subdwarfs is, in general, still unclear.
They can be formed only if the progenitor loses its envelope almost entirely after passing the red giant branch.
Different scenarios have been proposed to explain the huge mass loss.
There are three main such scenarios for hot subdwarf stars: The common envelope (CE) ejection \citep{1984ApJ...277..355W}, the stable Roche lobe overflow (RLOF) and the merger of double helium white dwarfs (HeWD) \citep{2002MNRAS.336..449H}.
Population synthesis studies show that the first two channels produce hot subdwarfs in binary systems and the last one creates single He-rich sdO stars \citep{2003MNRAS.341..669H, 2008A&A...484L..31H, 2012MNRAS.419..452Z}.
In the transition between the sdB and sdO types both the late hot-flasher scenario \citep{1996ApJ...466..359D,2004A&A...415..313M,2008A&A...491..253M} and the merger of helium white dwarfs with low mass main sequence stars \citep{2017ApJ...835..242Z} have been identified as a viable formation channel for some intermediate He-rich ($i$He) hot subdwarf stars.

Up to now, there is not a single formation theory that could simultaneously explain all the observed features and distributions of hot subdwarf stars.
With the advent of the Gaia survey \citep{2016A&A...595A...1G} and new spectroscopic surveys like LAMOST \citep{2012RAA....12.1197C} (the Large Sky Area Multi-Object Fiber Spectroscopic Telescope, also named the "Guo Shou Jing" Telescope),  we can characterize the details of hot subdwarfs by combining kinematic and spectroscopic properties of large observed samples.
By combing Gaia DR2 and LAMOST DR5, we presented the spectral analyses of 892 non-composite spectra hot subdwarf stars and the kinematics of 747 stars and revealed that hot subdwarfs can be further divided into four He groups that correlate with different formation channels \citep{2019ApJ...881....7L}.
Most recently, \cite{2020ApJ...898...64L} studied the kinematics of 182 single-lined hot subdwarfs selected from Gaia DR2 with spectra from LAMOST DR6 and DR7 \citep{2020ApJ...889..117L} and confirmed our results reported by \cite{2019ApJ...881....7L}.
Although 39\,800 hot subdwarf candidates were selected by \citet{2019A&A...621A..38G}, the properties of the majority of the stars are still unknown.

In this paper, a total of 1587 hot subdwarf stars are identified from the Gaia DR2 hot subdwarf candidate catalog \citep{2019A&A...621A..38G} with LAMOST DR7 spectra, among which, there are 224 new hot subdwarfs, which are confirmed here the first time.
We performed a spectral and kinematical analysis of all 1587 hot subdwarfs with LAMOST DR7 spectra using their Gaia EDR3 parallaxes and proper motions.
The catalogue of 1587 hot subdwarf stars includes RVs, atmospheric parameters, as well as Galactic space positions, space motions, orbital parameters and Galactic population classifications.
The paper is organized as follows:
Our data and sample selection are described in Sect 2.
The methods to derive atmospheric parameters, Galactic space velocities and orbital parameters are explained in Sect 3 and 4.
This is followed by our results on the properties of atmospheric parameters, Galactic space distributions, Galactic velocity distributions, Galactic orbits and Galactic population classifications in Sect 5.
Then we make a discussion in Sect 6, before summarizing and concluding our work in Sect 7.

\section{Data and sample selection} \label{sec:targ}
\subsection{Data}
LAMOST is a 4-m specially designed Schmidt spectroscopic survey telescope, which can simultaneously observe 4000 targets per exposure in a field of view of about $5^\circ$ in diameter \citep{2012RAA....12.1197C,2012RAA....12.1243L}.
LAMOST spectra are similar to SDSS data and cover the wavelength range of $3800-9100$ \AA\ with a resolution of $R\sim 1800$.
In March 2020, LAMOST has released 10\,608\,416 spectra in DR7, which is described and available at: \url{http://dr7.lamost.org/}

Gaia is a European Space Agency space telescope that maps the positions and motions of more than 1 billion stars to the highest precision yet of any missions \citep{2016A&A...595A...1G}. On April 25, 2018, Gaia DR2 was released, which provides high-precision positions ($\alpha$ and $\delta$), proper motions ($\mu_{\alpha}\cos\delta$ and $\mu_{\delta}$) and parallaxes ($\bar{\omega}$) as well as three broadband magnitudes ($G$, $G_{\rm BP}$ and $G_{\rm RP}$) for over 1.3 billion stars brighter than $G=21$ mag \citep{2018A&A...616A...1G}.
For the vast majority of stars in Gaia DR2, a reliable distance $D$ cannot be obtained by simply inverting the parallax, $D=1/\bar{\omega}$.
\cite{2018AJ....156...58B} presented a Gaia DR2 distance catalogue by estimating distances from parallaxes.
Gaia EDR3 \citep{2020arXiv201201533G} was released on December 3, 2020 and significantly improved the precision and accuracy of astrometry and broad-band photometry over Gaia DR2.
Gaia EDR3 contains astrometry and photometry for 1.8 billion sources brighter than $G=21$ mag.
For 1.5 billion of those sources, parallaxes ($\bar{\omega}$), proper motions ($\mu_{\alpha}\cos\delta$ and $\mu_{\delta}$) and the $G_{\rm BP}-G_{\rm RP}$ colours are available.
\cite{2021AJ....161..147B} also provided a Gaia-EDR3 distance catalogue based on Gaia EDR3.
Therefore, where necessary, unreliable distances were replaced with the estimated values from the Gaia-DR3 distance catalogue \citep{2021AJ....161..147B}.

\subsection{Sample selection}
\begin{figure*}
\epsscale{0.9}
\plotone{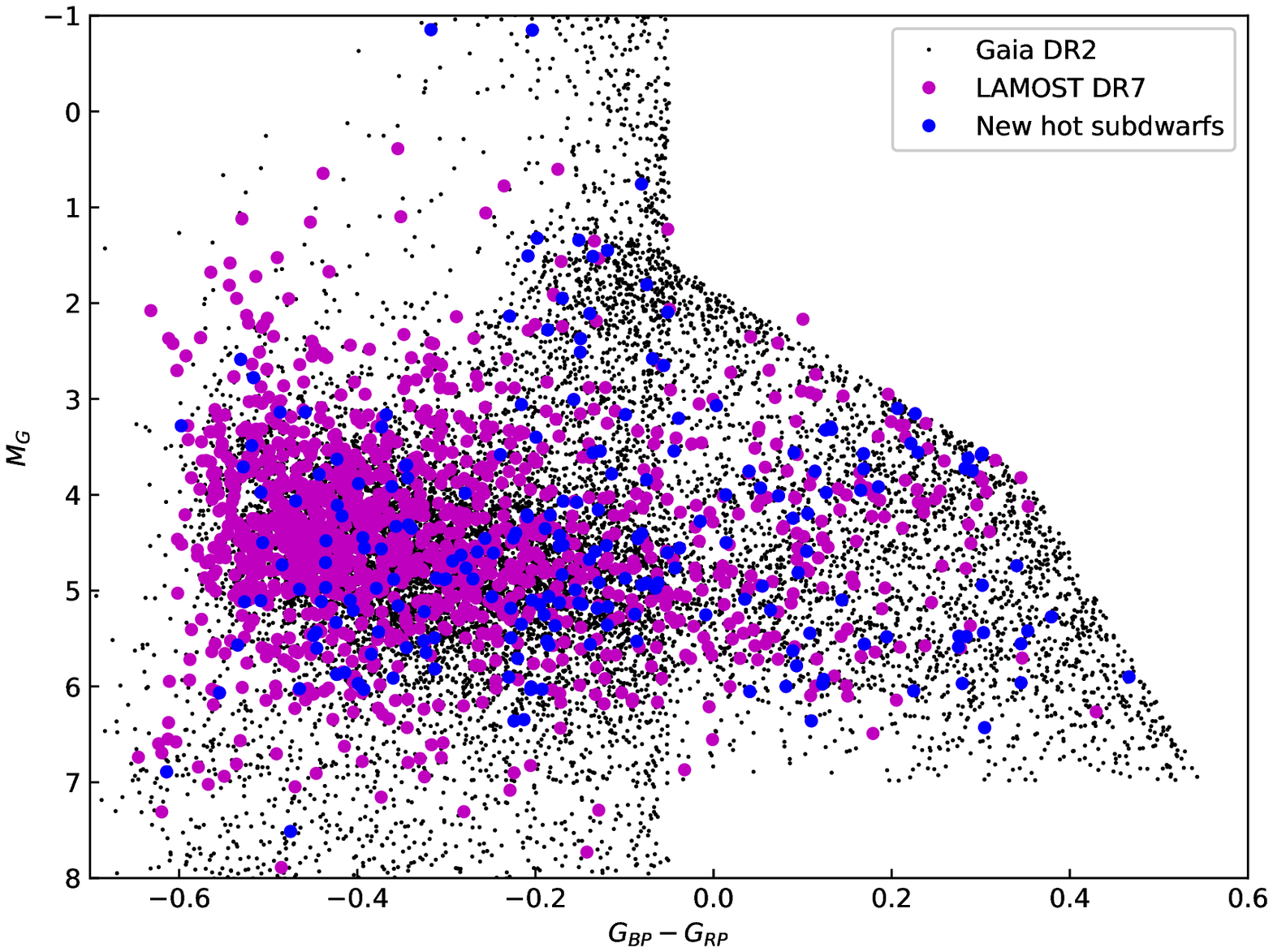}
\caption{Gaia Hertzsprung-Russell diagram of stars selected from the Gaia DR2 hot subdwarf candidate catalogue.
The magenta dots represent the 1363 known hot subdwarf stars observed by LAMOST DR7 and the blue ones denote the 224 newly discovered hot subdwarf stars form the Gaia DR2 hot subdwarf candidate catalogue with spectra of LAMOST DR7.\label{fig:fig1}}
\end{figure*}

Based on Gaia DR2 data, \citet{2019A&A...621A..38G} compiled an all-sky catalogue of 39\,800 hot subdwarf candidates by using the means of colour, absolute magnitude and reduced proper motion cuts.
\cite{2020ApJ...898...64L} showed that most of those objects lie between 500\,pc and 1500\,pc distance and the catalogue is nearly volume complete.
By cross-matching the catalogue with LAMOST DR7, we selected 2595 stars from the Gaia DR2 catalogue of hot subdwarf candidates.
After rejecting bad spectra, MS stars, WDs, and objects with strong Ca\,{\sc ii} H\&K  ($\lambda3933$ \AA\ and $\lambda3968$ \AA), Mg\,{\sc i} ($\lambda5183$ \AA), or Ca\,{\sc ii} ($\lambda8650$ \AA) absorption lines,
we obtained 1587 stars with a spectral signal-to-noise (S/N) over 10 in the $g$-band, by visually comparing them with reference spectra of hot subdwarf stars. 224 stars are confirmed as hot subdwarfs for the first time.
Figure\,\ref{fig:fig1} shows the positions of these 1587 stars in the Gaia HR diagram and Table \ref{tab:tab1} lists the proper motions and distances derived from Gaia EDR3 parallaxes.
However, for 74 stars reliable distances could not be calculated by inverting a negative parallax.
Therefore, we replaced their distances with the estimated values from the Gaia-EDR3 distance catalogue \citep{2021AJ....161..147B}. Additionally, Table \ref{tab:tab2} lists 665 targets removed from the Gaia DR2 catalogue of hot subdwarf candidates with LAMOST spectra with $S/N>10$ in the $g$-band.
Most of these candidates were misclassified as hot subdwarfs.

\begin{figure*}
\epsscale{0.9}
\plotone{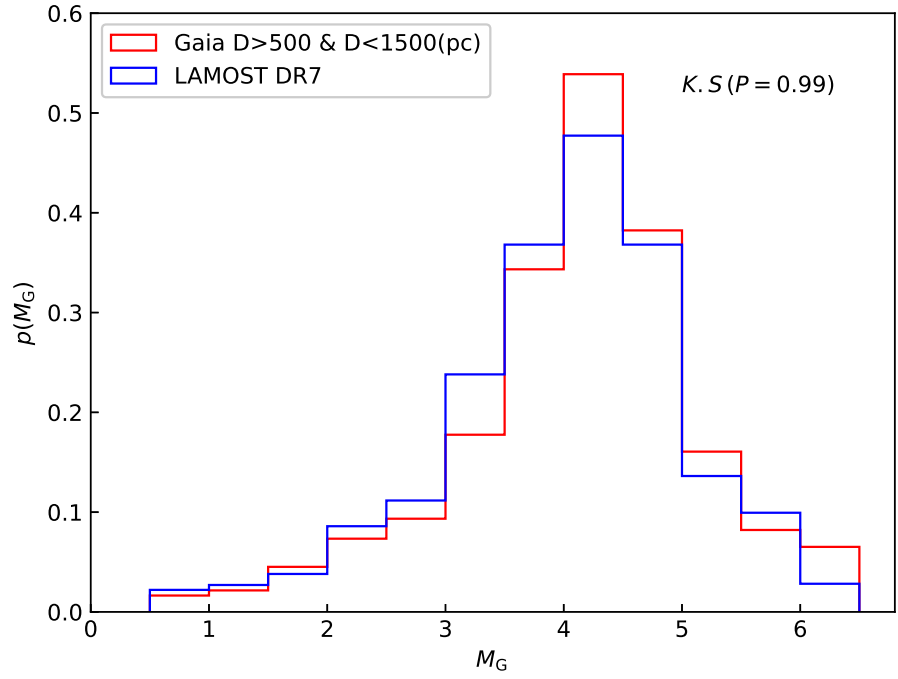}
\caption{Comparison of the distribution functions of the Gaia absolute magnitudes $M_{G}$ for the hot subdwarf stars in LAMOST DR7 and Gaia DR2 sample in the distance interval of $500<D<1\,500$ pc \citep{2019A&A...621A..38G}.
\citet{2020ApJ...898...64L} showed that the Gaia sample in the distance interval of $500<D<1\,500$ pc is volume complete. \label{fig:fig2}}
\end{figure*}

Figure\,\ref{fig:fig2} illustrates distribution functions of the absolute Gaia magnitudes $M_{\rm G}$ of hot subdwarf stars with a distance between 500\,pc and 1500\,pc in Gaia DR2 and LAMOST DR7.
The similar distributions suggest that the sample of hot subdwarfs in LAMOST DR7 is a representative and unbiased sub-sample of the volume-complete Gaia sample.
We consider all stars to be members of the thin disk, thick disk or halo populations until they can be further constrained.
Unknown and unresolved binary systems affect the calculations of Galactic velocities and orbits.
Although our study focuses on studying only single-lined hot subdwarf stars, based on a single epoch radial velocity measurement we cannot exclude the possibility of having unknown and unresolved binary systems in our sample.
Our simulations \citep{2020ApJ...898...64L} have demonstrated that the impact of an RV variability as a selection effect on the population classification propagates to less than 5\% relative differences in the number of stars in a He-group.

\section{Atmospheric parameters }
\subsection{Model atmospheres}
All model atmospheres were computed with the non-LTE (non-local thermodynamic equilibrium) code {\sc Tlusty} \citep{1995ApJ...439..875H}, version 205.
The corresponding hot subdwarf synthetic spectra were created using the spectrum synthesis code {\sc Synspec} \citep{2011ascl.soft09022H}, version 51.
Further details of these codes are available in the newest user's manuals \citep{2017arXiv170601859H,2017arXiv170601935H,2017arXiv170601937H}.
Model atmospheres were calculated under the standard assumptions of plane-parallel, horizontally homogeneous atmospheres in hydrostatic and radiative equilibrium.
Pure H+He chemical compositions were assumed for all models.
The broadening of H line profiles were calculated using the tables of \citet{2009ApJ...696.1755T} that take into account the effects of level dissolution directly in the evaluation of line profiles.
Four detailed He\,{\sc i} line profiles ($\lambda4026$, $\lambda4388$, $\lambda4471$ and $\lambda4922$ \AA) were taken from the special line broadening tables of \citet{1974ApJ...190..315M} and He\,{\sc ii} line profiles were taken from the Stark broadening tables of \citet{1989A&AS...78...51S}.

\subsection{Spectral analysis}
The radial velocities provided in the LAMOST catalog are not reliable for hot subdwarf stars, because hot subdwarfs are not included in the LAMOST stellar template library.
Therefore, we measured new radial velocities using our synthetic spectra as templates and list these velocities in Table \ref{tab:tab1}.
Following the works of \cite{2016ApJ...818..202L,2019ApJ...881....7L}, atmospheric parameters (effective temperature $T_{\rm eff}$, surface gravity $\log g$, and He abundances $y=n{\rm He}/n{\rm H}$) were obtained by fitting the observed data with a grid of synthetic spectra. The synthetic spectra were degraded to the resolution of LAMOST data and normalized in 80 \AA\ sections to the LAMOST flux-calibrated observations.
We used the wavelength range of $3800-7200$ \AA, which includes all significant H and He lines in the LAMOST spectra.
We employed the library for non-linear least-squares minimization and curve-fitting for Python ({\it LMFIT}) to determine the best-fit values of the parameters and estimate the standard errors.
We used the Levenberg-Marquardt minimization method of the {\it LMFIT} package in our fitting procedure.

\begin{figure*}
\epsscale{1.2}
\plotone{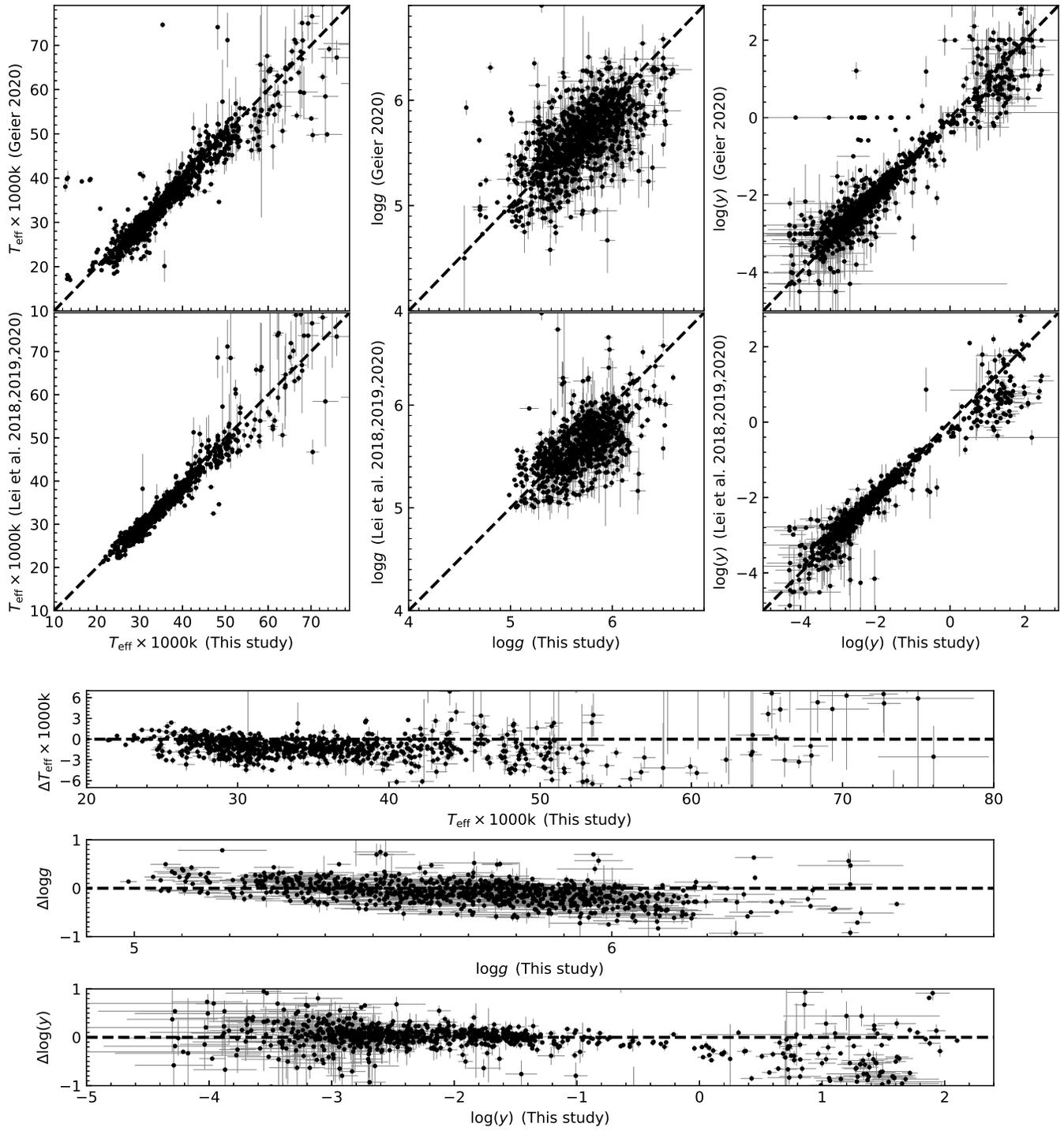}
\caption{Atmospheric parameters comparisons with the catalog of \citet{2020A&A...635A.193G} and the results of \citet{2018ApJ...868...70L, 2019ApJ...881..135L,2020ApJ...889..117L}. The bottom three panels show the residuals between the results of this study and the ones of \citet{2018ApJ...868...70L, 2019ApJ...881..135L,2020ApJ...889..117L}\label{fig:fig3}}
\end{figure*}

Table 1 lists the atmospheric parameters derived from the spectra of the 1587 hot subdwarf stars.
We cross-matched our results with the latest catalogue of known hot subdwarf stars \citep{2020A&A...635A.193G} and obtained 1363 stars in common. $T_{\rm eff}$, $\log{g}$, and $\log{(y)}$ are simultaneously available for 1195 common stars.
The top three panels of Figure\,\ref{fig:fig3} show comparisons of the atmospheric parameters of these stars between our study and \citet{2020A&A...635A.193G}.
The derived $T_{\rm eff}$ and $\log{(y)}$ are similar in both of these studies.
The largest deviations appear above 55\,000 K, where both non-LTE effects and model composition play an important role.
For the He abundance the largest errors are random errors at the lowest and highest He abundances.
Surface gravity is different, though.
Instead of a linear correlation, it shows an elongated blob.
The surface gravity is particularly difficult to measure with precision from H+He models and low resolution observations, and this affects both studies in our comparison.
Although the $\log{g}$ values show a larger dispersion than the former two parameters, our results agree with the values from \citet{2020A&A...635A.193G} and the error bars are also comparable.
We note that our error bars are 1$\sigma$ errors (68\% confidence), while the collection of \citet{2020A&A...635A.193G} also includes samples calculated for 60\% confidence.
There are 852 stars in the sample of \citet{2018ApJ...868...70L, 2019ApJ...881..135L, 2020ApJ...889..117L}, which were analysed by applying the same model atmospheres ({\sc Tlusty}) on the same spectra, but a different fitting procedure. The bottom three panels of Figure\,\ref{fig:fig3} show the atmospheric parameter comparisons between our study with \citet{2018ApJ...868...70L, 2019ApJ...881..135L, 2020ApJ...889..117L}. $T_{\rm eff}$ is quite similar for these stars. $\log{(y)}$ agrees with results of \citet{2018ApJ...868...70L, 2019ApJ...881..135L, 2020ApJ...889..117L} for stars with $\log{(y)}<0$ while it shows a linear offset for stars with $\log{(y)}>0$.
As shown in above results, $\log{g}$ shows exhibits a larger dispersion than the former two parameters.
It is reassuring that the largest systematic differences are observed at hot, He dominated stars, which require more complex models.

\section{Calculations of Galactic space velocities and orbital parameters}\label{sec:cal}
Combining the parallaxes and proper motions published by Gaia EDR3 and the RV measured from the spectra of LAMOST DR7, we calculated the space velocity components of hot subdwarf stars in the right-handed Galactocentric Cartesian coordinate system with the \textit{Astropy} Python package.
We took the usual convention of the space velocity components $U$, $V$, and $W$ oriented towards
the Galactic center, the direction of Galactic rotation and the north Galactic pole, respectively.
We adopted the distance of the Sun from the Galactic center to be 8.4 kpc and the velocity of the local standard of rest (LSR) to be $242\,{\rm km\,s^{-1}}$ \citep{2013A&A...549A.137I}.
For the solar velocity components with respect to the LSR, we took ($U_{\odot}$, $V_{\odot}$, $W_{\odot}$)$=$(11.1, 12.24, 7.25)$\,{\rm km\,s^{-1}}$ \citep{2010MNRAS.403.1829S}.
Using \textit{Astropy}, we also computed the space position components ($X$, $Y$, $Z$) in a right-handed Galactocentric Cartesian reference frame.

The Galactic orbits were computed using the \textit{Galpy} Python package \citep{2015ApJS..216...29B}.
We integrated orbits in \textit{Galpy} "MWpotential2014" potential that comprises a power-law bulge with an exponential cut-off, an exponential disk and a power-law halo component \citep{2015ApJS..216...29B}, using the same Galactocentric distance of the Sun and the same LSR as in \textit{Astropy}.
The time of orbital integration was set from 0 to $3.5$\,Gyr in steps of $1$\,Myr.
The Galactic orbital parameters of hot subdwarf stars, such as the apocentre ($R_{\rm ap}$), pericentre ($R_{\rm peri}$), eccentricity ($e$), maximum vertical amplitude ($z_{\rm max}$), normalised z-extent ($z_{n}$) and z-component of the angular momentum ($J_{z}$), were extracted from integrating their orbital paths.
$R_{\rm ap}$ and $R_{\rm peri}$ denote the maximum and minimum distances of an orbit from the Galactic center, respectively.
We defined the eccentricity by
\begin{equation}
e=\frac{R_{\rm ap}-R_{\rm peri}}{R_{\rm ap}+R_{\rm peri}},
\end{equation}
and the normalised z-extent by
\begin{equation}
z_{n}=\frac{z_{\rm max}}{R(z_{\rm max})},
\end{equation}
where $R=\sqrt{X^{2}+Y^{2}}$ is the Galactocentric distance.

The space positions and velocity components, as well as the orbital parameters are listed in Table \ref{tab:tab1}.
The errors of these parameters were estimated with a Monte Carlo (MC) method. For each star, 1000 random input values with a Gaussian distribution were simultaneously performed and the output parameters were computed together with their errors. Further details on the parameter error calculations were presented by \citet{2019ApJ...881....7L}.

\section{Results} \label{sec:res}
\subsection{The atmospheric properties of the stars}
Figure\,\ref{fig:fig5} exhibits the distribution of hot subdwarf stars in the $T_{\rm eff}-\log\,g$ diagram. The zero-age HB (ZAHB) and terminal-age HB (TAHB) calculated by \citet{1993ApJ...419..596D} and the zero-age He main sequence given by \citet{1971AcA....21....1P} are marked in the figure. Three sdB evolutionary tracks of \citet{1993ApJ...419..596D} for solar metallicity and a subdwarf core mass of $0.47M_{\odot}$ are plotted in Figure\,\ref{fig:fig5}.

\begin{figure*}
\epsscale{0.9}
\plotone{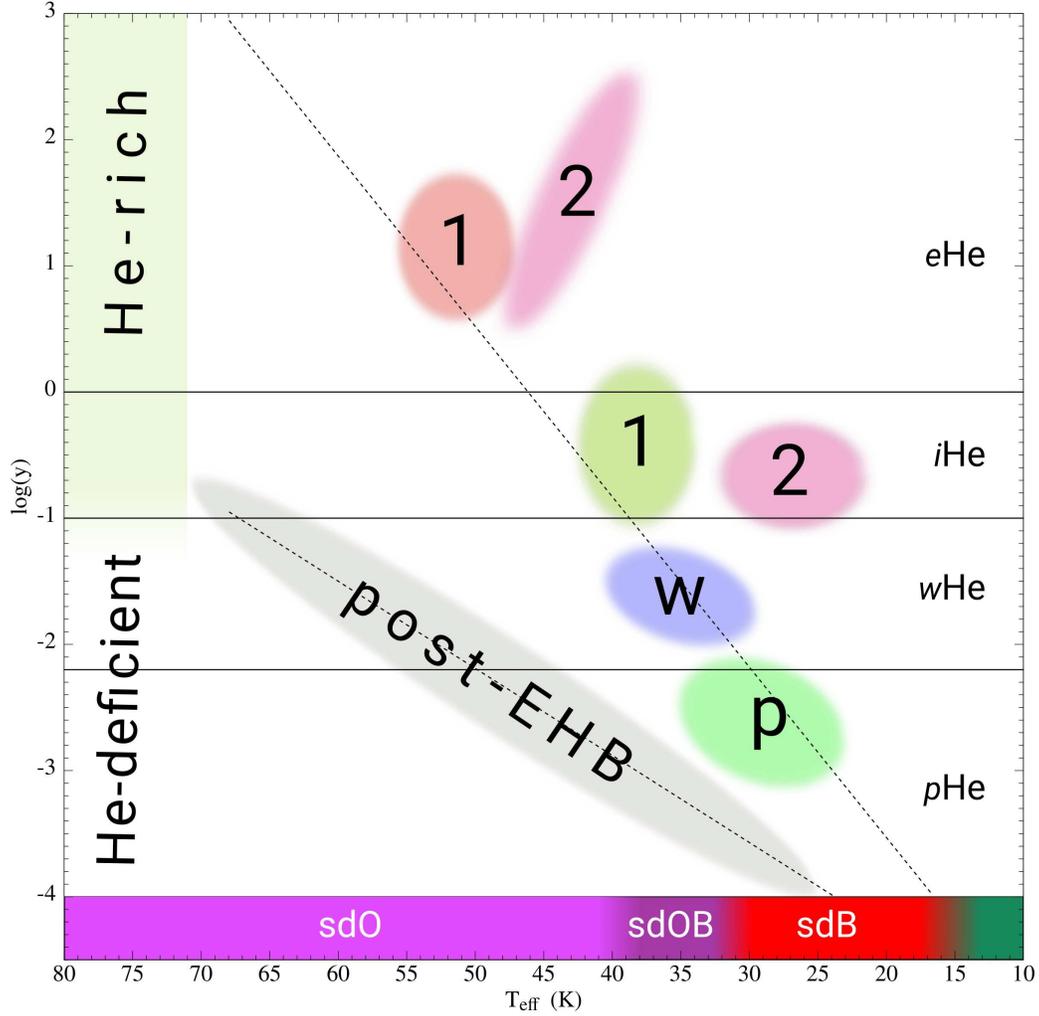}
\caption{
$T_{\rm eff}-\log{(y)}$ diagram for the He abundance classification scheme used in this paper.
The hot subdwarf parameter space is divided into He-rich and He-deficient stars, which are further subdivided to He-poor, He-weak, intermediate He-rich ($i$He-rich) and extreme He-rich ($e$He-rich) classes.
We find that both classes of He-rich stars have (at least) two groups of stars that separate in their atmospheric and kinematic properties.
These groups are outlined by four ellipses.
The dashed lines outline the $T_{\rm eff}-\log{(y)}$ correlation found by \citet{2003A&A...400..939E} for sdB stars and a similar correlation for post-EHB stars by \citet{2012MNRAS.427.2180N}.
Along these two correlations He-deficient stars separate into EHB (w and p) and post-EHB stars, which are evolutionary connected.
He-deficient sdB and sdOB stars on the EHB separate further into He-weak and He-poor groups, which are potential $p$ and $g$-mode pulsator candidates, respectively.
Although not all stars in these regions prove to be pulsating stars, while in between these groups hybrid pulsators were found showing both modes.
\label{fig:fig4}}
\end{figure*}

\begin{figure*}
\epsscale{0.9}
\plotone{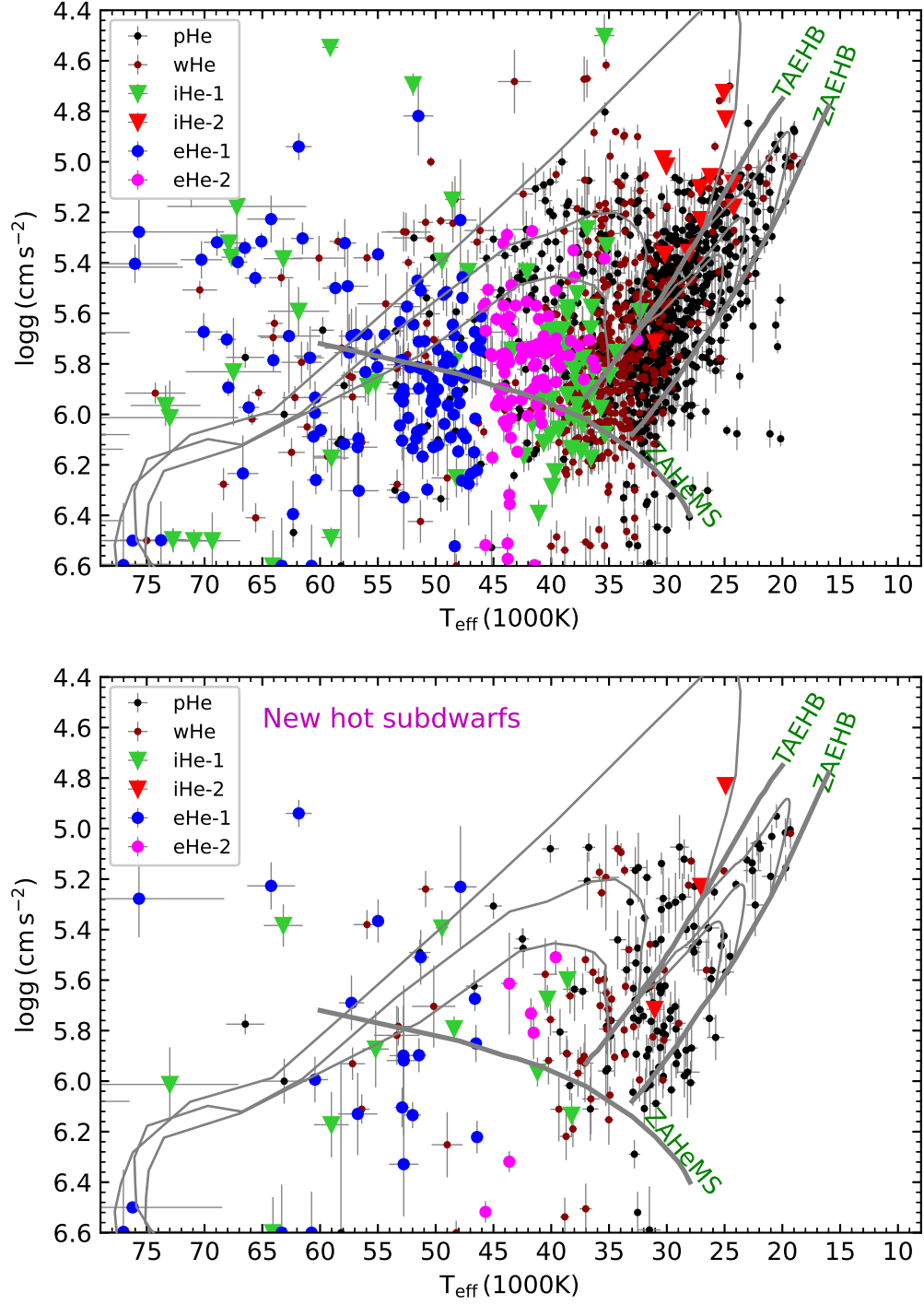}
\caption{$T_{\rm eff}-\log{g}$ diagram for hot subdwarf stars. The zero-age EHB (ZAEHB), terminal-age EHB (TAEHB) \citep{1993ApJ...419..596D}, and zero-age He main sequence (ZAHeMS) \citep{1971AcA....21....1P} are marked with the gray thick solid lines, respectively. The three gray thin solid lines express the evolutionary tracks of \cite{1993ApJ...419..596D} for solar metallicity and subdwarf masses from top to bottom: 0.480, 0.473 and $0.471\,{\rm M}_{\odot}$.  \label{fig:fig5}}
\end{figure*}

Following our helium abundance classification scheme, hot subdwarf stars are divided into He-rich and He-deficient stars with respect to the solar He abundance $\log{(y)}=-1$.
Both He-rich and He-deficient stars are then further divided into two classes (He abundance classes) via $\log{(y)}=0$ and $\log{(y)}=-2.2$, respectively.
As described by \citet{2012MNRAS.427.2180N} and \citet{2019ApJ...881....7L}, our classification scheme associates these classes with distinct formation channels in the $T_{\rm eff}-\log{g}$ diagram.
In this paper, He-rich stars with $\log{(y)} \ge 0$ are marked as the extreme He-rich ($e$He) stars while He-rich stars with $-1\le \log{(y)}<0$ are named as intermediate He-rich ($i$He) stars.
We refer to He-deficient stars with $-2.2\le\log{(y)}<-1$ as He-weak ($w$He) and $\log{(y)}<-2.2$ stars as He-poor ($p$He) stars.
Within the four He abundance classes, we distinguish groups of stars that show characteristically different atmospheric and kinematic properties.
Figure\,\ref{fig:fig4} gives a schematic overview of the He abundance classification system.

In Figure\,\ref{fig:fig5}, four classes of hot subdwarf stars can be outlined by using our He abundance classification scheme.
As reported in previous studies \citep{2012MNRAS.427.2180N, 2016ApJ...818..202L, 2019ApJ...881....7L}, two He-deficient sdB groups are located on the EHB and they correspond to potential $g-$mode and $p-$mode pulsating sdB stars.
However, the correspondence between pulsation and atmospheric properties are not yet clear and only a tiny fraction of He-deficient stars are pulsators, therefore we distinguish these groups by the He abundance.
One group consists of He-poor  ($p$He) stars with $\log{(y)}<-2.2$ near $T_{\rm eff}=29\,000\,$K and $\log{g}=5.5$\,cm\,s$^{-2}$.
Another group consists of He-weak ($w$He) stars with $-2.2\le\log{(y)}<-1$ near $T_{\rm eff}=34\,000\,$K and $\log{g}=5.9$\,cm\,s$^{-2}$.
The former group is on average 10 times less He abundant than the latter.
He-rich stars ($\log{(y)}>-1$) also show two classes at higher He abundances.
One class is composed of $e$He-sdO/sdB stars ($\log{(y})\ge0$) between $T_{\rm eff}=36\,000\,$K and $T_{\rm eff}=57\,000\,$K and another class consists of the $i$He-rich sdO/sdB stars ($-1\le\log{(y)}<0$) near $T_{\rm eff}=38\,000\,$K and $\log{g}=5.9$\,cm\,s$^{-2}$.
These results are in agreement with the observations of \cite{2012MNRAS.427.2180N} and \cite{2016ApJ...818..202L, 2019ApJ...881....7L}.

\begin{figure*}
\epsscale{1.2}
\plotone{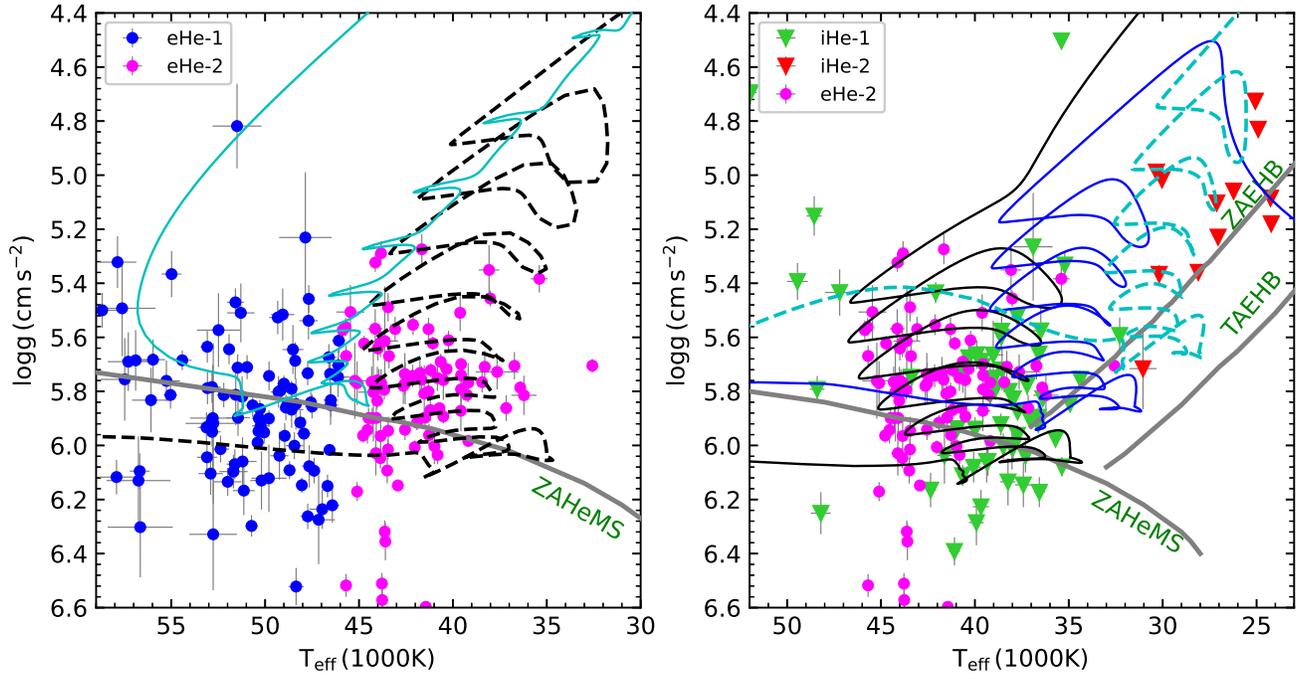}
\caption{Left:$T_{\rm eff}-\log{g}$ diagram for the double WD merger channels.  The cyan solid and black dashed curves express the evolutionary tracks for a hot subdwarf mass of 0.8 and $0.5\,{\rm M}_{\odot}$ through the double WD merger channels of \cite{2012MNRAS.419..452Z}.  Right: $T_{\rm eff}-\log\,g $ diagram for the hot-flasher scenario.
The cyan dashed curve denotes the evolutionary track for a hot subdwarf mass of $0.47426\,{\rm M}_{\odot}$ through the hot-flasher scenario with no He enrichment, the blue solid curve for a subdwarf mass of $0.47378\,{\rm M}_{\odot}$ with shallow mixing (SM), and the black solid curve for a hot subdwarf mass of $0.47112\,{\rm M}_{\odot}$ with deep mixing from \citet{2008A&A...491..253M}.\label{fig:fig6}}
\end{figure*}

The left panel of Figure\,\ref{fig:fig6} illustrates the distribution of $e$He-rich hot subdwarf stars in the $T_{\rm eff}-\log\,g$ diagram.
Two evolutionary tracks with solar metallicity for subdwarf masses of $0.5$ and $0.8M_{\odot}$ from the double WD merger channels of \citet{2012MNRAS.419..452Z} are also marked.
It is clearly seen that $e$He-rich stars with $\log{(y)}\ge0$ show a gap near $T_{\rm eff}=46\,000$\,K in the $T_{\rm eff}-\log{g}$ diagram, which were reported in several studies \citep{2007A&A...462..269S, 2009PhDT.......273H, 2019MNRAS.486.2169K}.
Two groups ($e$He-1 and $e$He-2) of $e$He-rich stars with $\log(y)\ge0$ can be outlined by the gap.
They cluster either near $T_{\rm eff}=43\,000$\,K and $\log{g}=5.8$\,cm\,s$^{-2}$, or near $T_{\rm eff}=51\,000$\,K and $\log{g}=6.0$\,cm\,s$^{-2}$, respectively.
When compared to evolutionary tracks of hot subdwarf stars through the double WD merger model \citep{2012MNRAS.419..452Z},  the evolutionary tracks match these two groups.
Therefore, Group $e$He-1 we associated with the fast merger channel and Group $e$He-2 with the slow merger channel in the merger model of \cite{2012MNRAS.419..452Z}.

At higher temperature, the majority of $i$He-rich stars with $-1\le\log{(y)}<0$ cluster near $T_{\rm eff}=38\,000$\,K and $\log{g}=5.9$\,cm\,s$^{-2}$ and define Group $i$He-1.
At lower temperature, we find 12 $i$He-rich stars with $-1\le\log{(y)}<0$, near $T_{\rm eff}=27\,000$\,K and $\log{g}=5.2$\,cm\,s$^{-2}$ in the $T_{\rm eff}-\log{g}$ diagram and outlines the newly discovered Group $i$He-2.
Although $i$He-rich stars with $-1\le\log{(y)}<0$ and $T_{\rm eff}<32\,000$\,K did not show up as a clear group in LAMOST DR5, one can see that in the current, and larger sample, 5 stars are found in the region of Group $i$He-2.
The right panel of Figure\,\ref{fig:fig6} compares evolutionary tracks for hot subdwarfs via the hot-flasher scenario \citep{2008A&A...491..253M} for three stellar surface mixing: Hot-flasher with no He enrichment, hot-flasher with shallow mixing (SM), and hot-flasher with deep mixing (DM). It appears that these evolutionary tracks not only match Group $i$He-1 and $i$He-2 at lower temperatures, but also cover Group $e$He-2 at higher temperature.

\begin{figure*}
\epsscale{0.9}
\plotone{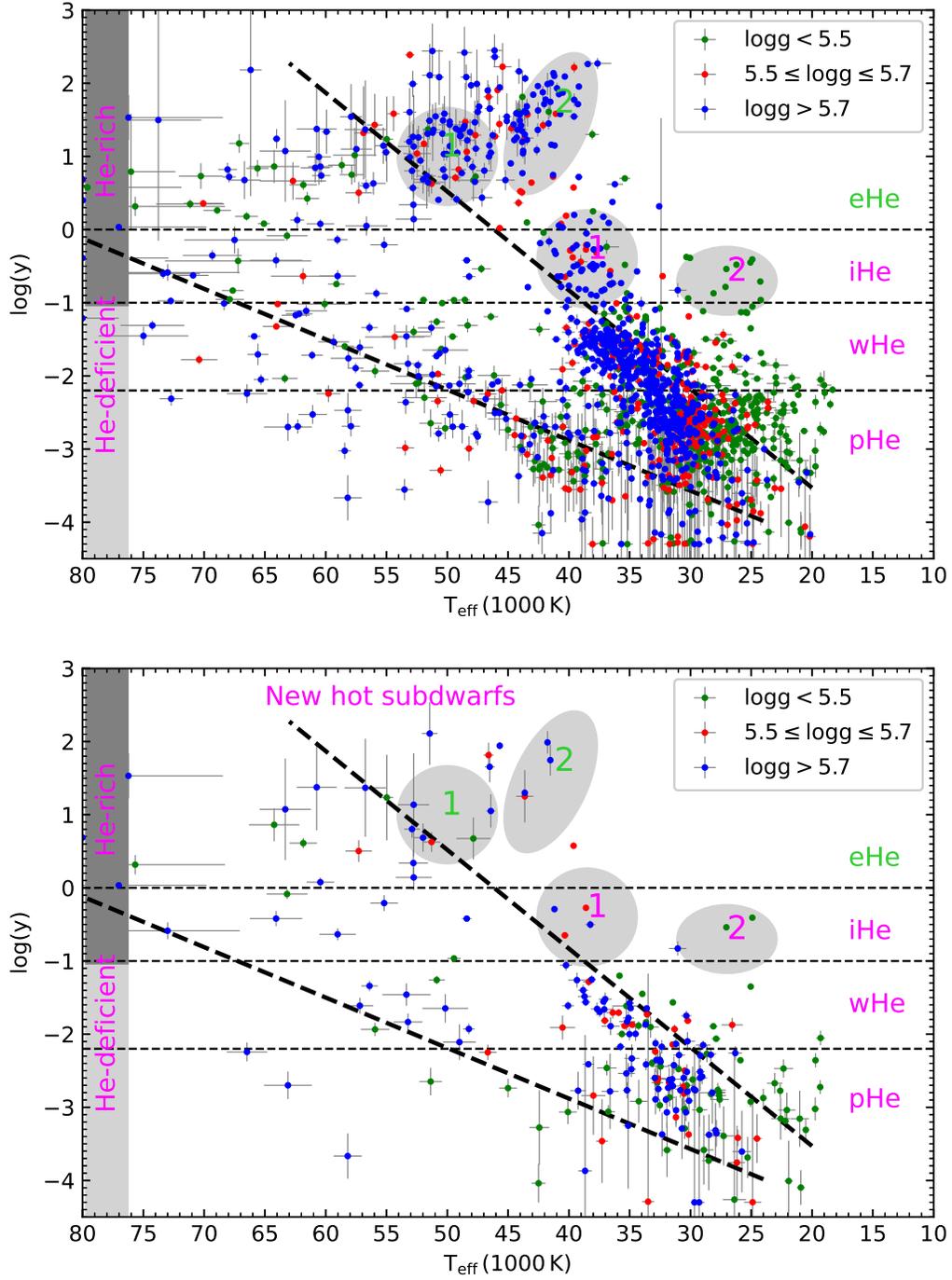}
\caption{Helium abundance versus effective temperature.
The thick dashed lines represent the best fitting trends, the top one from \cite{2003A&A...400..939E} for EHB stars, and the bottom one from \cite{2012MNRAS.427.2180N} for post-EHB stars.
Three thin dashed lines denote $\log(y)=0$, $\log(y)=-1$, and $\log(y)=-2.2$, which are notable gaps in He abundance and the basis of our He abundance classification scheme.
The grey ellipses outline four groups of He-rich stars, that also separate in their kinematic properties.\label{fig:fig7}}
\end{figure*}

Figure\,\ref{fig:fig7} displays the distribution of hot subdwarf stars in the $T_{\rm eff}-\log{(y)}$ diagram.
As reported by other authors \citep{2003A&A...400..939E, 2012MNRAS.427.2180N, 2019ApJ...881....7L}, our sample also shows two distinct He sequences with a clear trend of increasing He abundance with effective temperature.
To make a comparison with previous results we plot the two best-fit trends in Figure\,\ref{fig:fig7}, that are described by the following relationships:
\begin{equation}
\rm{I:}
\log(y)=-3.53+1.35(T_{\rm{eff}}/10^{4}\rm{K}-2.00),
\end{equation}
\begin{equation}
\rm{II:}
 \log(y)=-4.26+0.69(T_{\rm{eff}}/10^{4}\rm{K}-2.00).
\end{equation}
Where the former is taken from \citet{2003A&A...400..939E} and the latter is from \citet{2012MNRAS.427.2180N}.
The first best-fit trend is able to match He-deficient stars in the first sequence, but it is not suitable for He-rich stars.
As reported by other authors \citep{2009PhDT.......273H, 2012MNRAS.427.2180N}, He-rich stars with $\log{(y)}\ge0$ follow an opposite trend: The He abundance decreases with temperature and approaches $\log{(y)}=-0.5$, and $i$He-rich stars with $-1\le\log{(y)}<0$ deviate progressively from the best-fit trend with increasing temperature.
Moreover, the four groups of He-rich stars identified in the $T_{\rm eff}-\log{g}$ diagram can be clearly outlined in the $T_{\rm eff}-\log{(y)}$ diagram as well.
They lie near or above the first sequence (Eqn 3) and are clearly marked by four ellipses in the $T_{\rm eff}-\log{(y)}$ diagram.
Form the $T_{\rm eff}-\log{(y)}$ diagram, one can see that 11 stars with $0\le\log{(y)}<0.25$ are near Group $i$He-1, and 5 stars with $-1.14\le\log{(y)}<-1$ are near Group $i$He-2.
They are outlined as Group $i$He-1 and $i$He-2, respectively.
He-rich stars aggregate in four regions in the $T_{\rm eff}-\log{g}$ and $T_{\rm eff}-\log{(y)}$ diagrams.
As shown in Figure\,\ref{fig:fig7}, the four groups can be defined by the four ellipses and they are well separated by the observed gaps in the $T_{\rm eff}-\log{(y)}$ diagram.
Groups $e$He-1 and $e$He-2 show significantly different trends in He abundance: with the increase of temperature, the He abundance of Group $e$He-2 declines linearly to the value of Group $e$He-1.
The He abundance of Group $e$He-1 is nearly independent of temperature.
A similar result was reported by \cite{2021MNRAS.501..623J}.
The number of stars in Group $i$He-1 and $i$He-2 display a significant difference: the former group has 65 members, but the later includes only 17 stars.
The He abundance of Group $i$He-1 is on average somewhat higher than that of Group $i$He-2.
\subsection{Space distribution} \label{subsec:spac}

\begin{figure*}
\epsscale{1.0}
\plotone{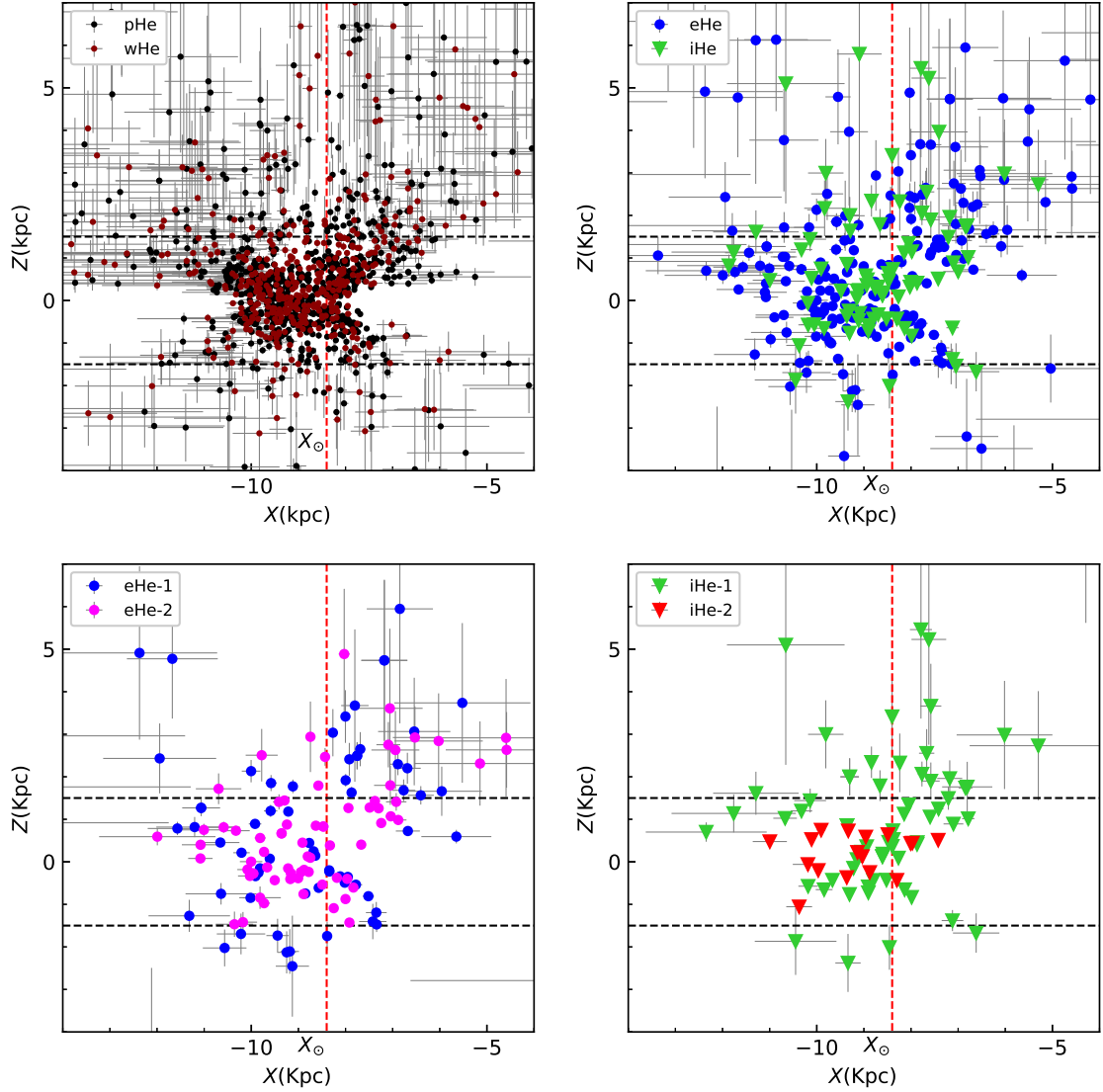}
\caption{The distribution of the hot subdwarf stars in the Cartesian Galactic $X-Z$ coordinates.
He-deficient stars are shown in the upper left panel and He-rich stars are displayed in the upper right panel.
Groups $e$He-1 and $e$He-2 are exhibited in the lower left panel and Groups $i$He-1 and $i$He-2 are illustrated in the lower right panel.
The red dashed line marks the solar position and the black dashed lines express $Z=\pm1\,500$\,pc.
\label{fig:fig8}}
\end{figure*}

Figure\,\ref{fig:fig8} exhibits the space positions of the four hot subdwarf helium abundance classes and four groups in the $X-Z$ diagrams, in which the black dashed lines mark $Z=\pm1.5$\,kpc, which corresponds to the vertical scale height of the thick disk \citep{2017MNRAS.467.2430M}.
The upper left panel of Figure\,\ref{fig:fig8} displays that two He-deficient groups have nearly the same space position distributions.
 About $74\pm1\%$ of the stars lie within $|Z|<1.5$\,kpc.
Based on the upper right panel of Figure\,\ref{fig:fig8}, the two He-rich groups do not display any clear differences in space position distribution and about $63\pm3\%$ the stars lie within $|Z|<1.5$\,kpc.
However, we find $11\pm3\%$ difference between the relative number of He-rich and He-deficient stars (upper left and right panels) with $|Z|<1.5$\,kpc, which implies that He-rich and He-deficient hot subdwarfs have different kinematic origins.

The lower left panel of Figure\,\ref{fig:fig8} shows that $49\pm6\%$ of the stars in Group $e$He-1 are found at $|Z|<1.5$\,kpc, but Group $e$He-2 has about $78\pm5\%$ of the stars in this region.
The lower \textcolor{blue}{right} panel of Figure\,\ref{fig:fig8} reveals that $62\pm6\%$ of the stars in Group $i$He-1 are located at $|Z|<1.5$\,kpc, while in Group $i$He-2 all stars are in this region.
These observations suggest that all four groups of He-rich stars may have different kinematic origins or formation channels.

\subsection{Galactic velocity distribution} \label{subsec:cal}
\begin{figure*}
\epsscale{1.2}
\plotone{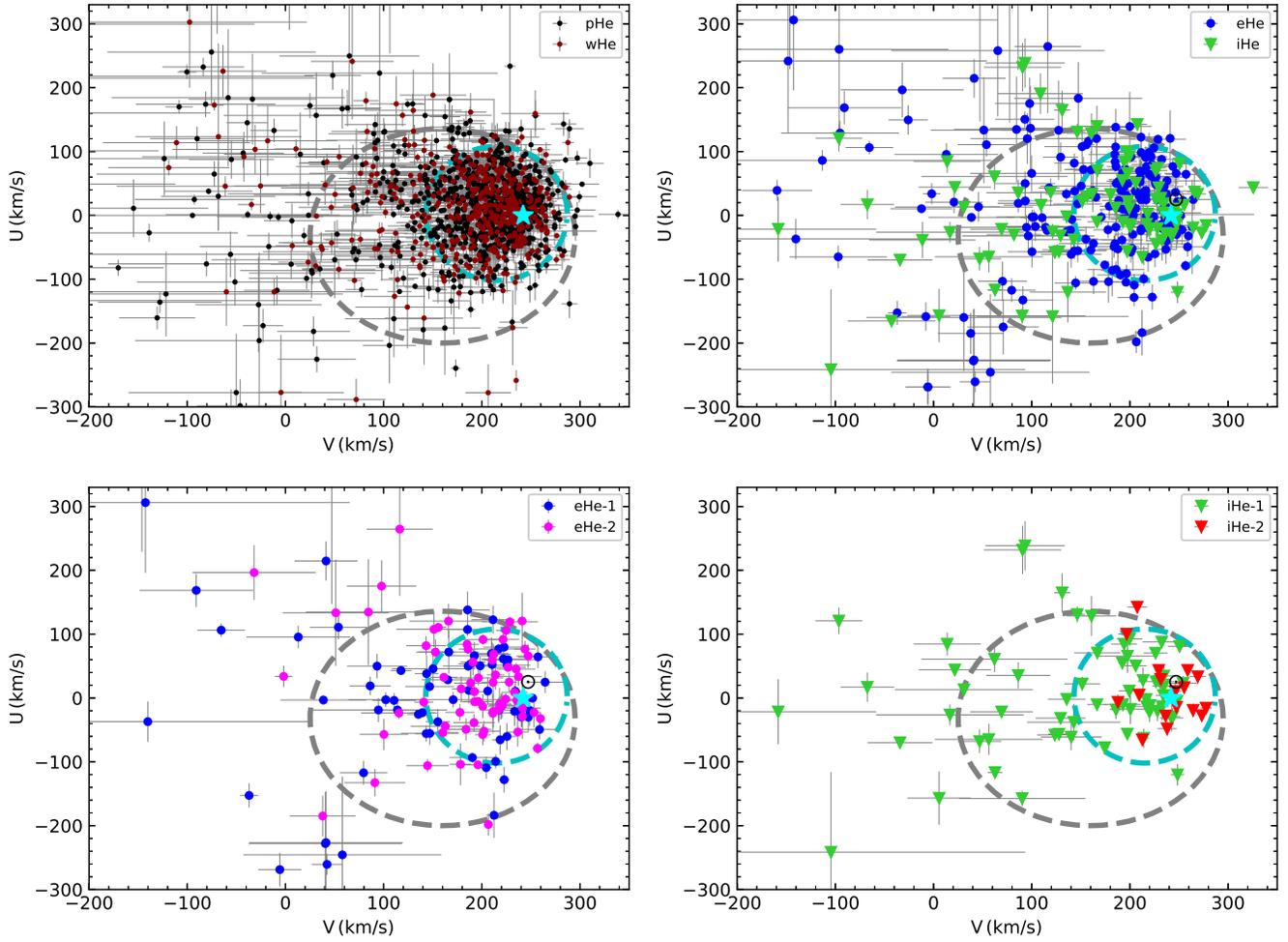}
\caption{$U-V$ velocity diagram for the four hot subdwarf helium abundance classes and four He-rich groups.
The two dashed ellipses denote the $3\sigma$ limits for the thin disk and thick disk populations from \citet{2006A&A...447..173P}, respectively.
The cyan star symbol represents the Local Standard of Rest (LSR).
\label{fig:fig9}}
\end{figure*}

Figure\,\ref{fig:fig9} displays the distribution of the four hot subdwarf helium classes and four He-rich groups in the $U-V$ velocity diagram.
In order to understand their kinematic origins, we also plot the two dashed ellipses as presented in Figure\,1 of \cite{2017MNRAS.467...68M}.
They represent the $3\sigma-$limits of thin and thick disk WDs \citep{2006A&A...447..173P}, respectively.

The upper left panel of Figure\,\ref{fig:fig9} exhibits that $70\pm1\%$ of He-deficient stars are located within the $3\sigma-$limits of the thin-disk, $18\pm1\%$ lie between the $3\sigma-$limits of thin and thick-disk and the remaining $12\pm1\%$ are located outside the $3\sigma-$limits of thick-disk stars.
The upper right panel of Figure\,\ref{fig:fig9} displays that the fractions of He-rich stars in the three regions defined above are $57\pm3\%$, $21\pm2\%$ and $21\pm2\%$, respectively.
Comparisons between the upper panels of Figure\,\ref{fig:fig9} demonstrate that the fraction of He-rich stars in the halo population is $9\pm2\%$ higher than that of He-deficient stars, which agrees with the results we found from the space distribution of stars in the preceding section.
These observations support that He-rich and He-deficient stars have different kinematic origins.

The lower left panel of Figure\,\ref{fig:fig9} shows a comparison between Group $e$He-1 and $e$He-2 of $e$He-rich stars with $\log{(y)}\geq0$.
There are $51\pm6\%$ of stars in Group $e$He-1 within the $3\sigma-$limits of the thin-disk, $27\pm5\%$ between the $3\sigma-$limits of the thin and thick disk, and the remaining $22\pm5\%$ are out of the $3\sigma-$limits of the thick-disk.
But Group $e$He-2 shows a quite different distribution, where $68\pm6\%$, $17\pm4\%$ and $15\pm4\%$ correspond to these three regions, respectively.
It can be concluded that stars in Group $e$He-1 and $e$He-2 mainly form in the thick and thin disk, respectively.
From the lower right panel of Figure\,\ref{fig:fig8}, one can see that $54\pm6\%$, $26\pm5\%$, $20\pm5\%$ stars in Group $i$He-1 are located above three corresponding regions and almost all stars ($94\pm6\%$) of Group $i$He-2 lie within the $3\sigma-$limits of the thin-disk.
Again, these results support that the four He-rich groups have different kinematic origins.

Table$\,$\ref{tab:tab3} lists the mean values and dispersions of the Galactic velocities of the four He classes and four He-rich groups.
Intermediate $i$He-rich stars with $-1\le\log{(y)}<0$ show the largest velocity dispersion values and $e$He-rich stars with $\log(y)\ge0$ exhibit the second largest velocity dispersion.
This result is consistent with the findings of \cite{2017MNRAS.467...68M} but in disagreement with our previous results \citep{2019ApJ...881....7L, 2020ApJ...898...64L}.
Thanks to the significant improvements in the astrometric precision and accuracy of EDR3, more $i$He-rich stars with $-1\le\log{(y)}<0$ are identified now as halo stars.
The two groups of He-deficient sdB stars show similar dispersions, and their mean values are less than those of He-rich stars, which further supports that He-deficient and He-rich stars have different kinematic origins.
The velocity dispersion of Group $e$He-1 is larger than of Group $e$He-2, but nearly equal to Group $i$He-1.
Group $i$He-2 exhibits the smallest velocity dispersion.
It can be concluded that the diverse range of kinematic velocities supports the different origins of the four He-rich groups.

\subsection{Galactic orbits} \label{sec:orb}
\begin{figure*}
\epsscale{1.2}
\plotone{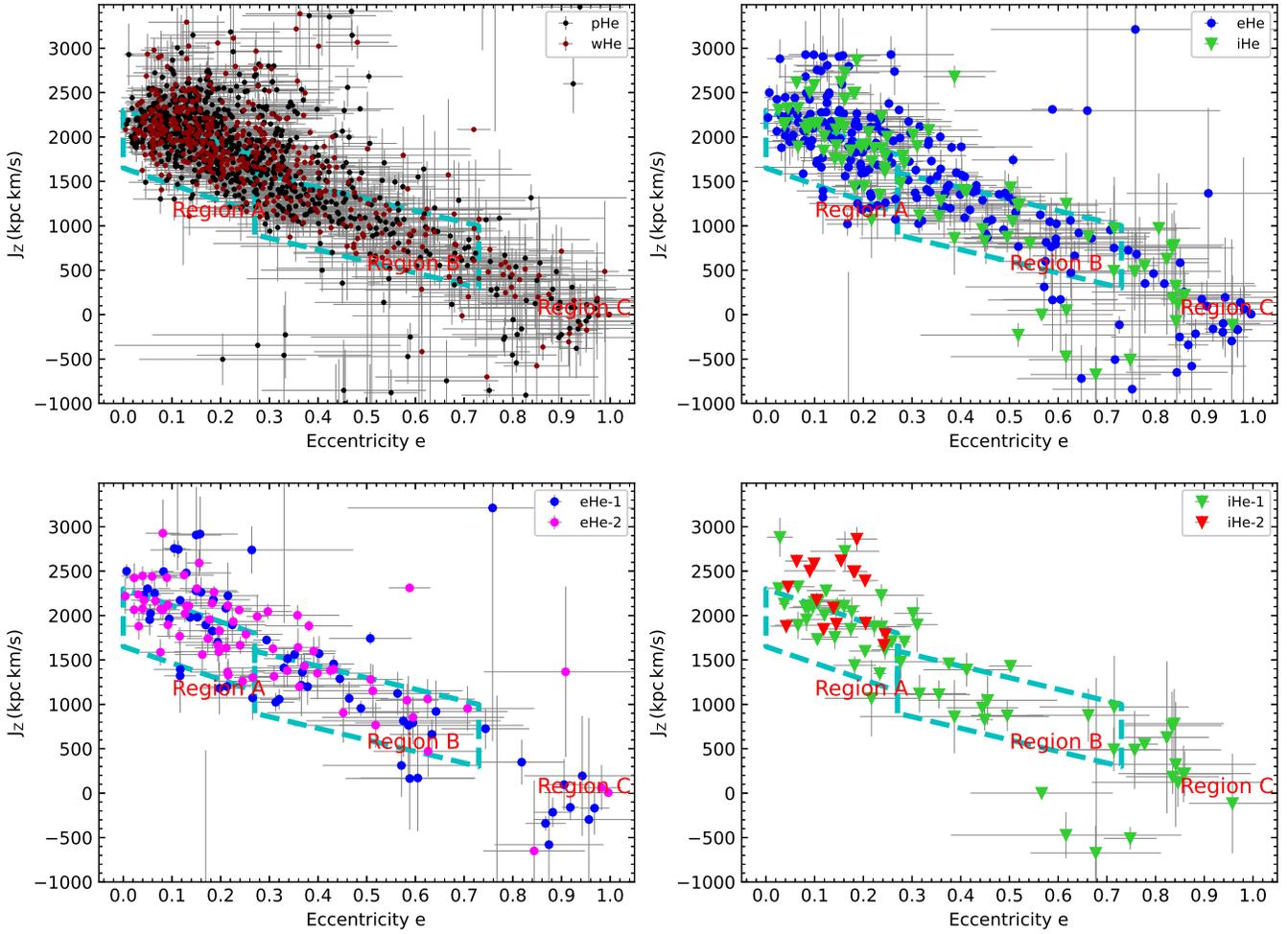}
\caption{The Z-component of the angular momentum ($J_{\rm z}$) versus eccentricity ($e$) for the four hot subdwarf helium abundance classes and four He-rich groups.
The two parallelograms denote Region A (thin disk) and Region B (thick disk) from \citet{2006A&A...447..173P}.
\label{fig:fig10}}
\end{figure*}

Both the z-component of the angular momentum $J_{z}$ and the eccentricity $e$ of the orbit are two important  orbital parameters for characterising the kinematics of hot subdwarf stars.
Figure\,\ref{fig:fig10} shows the distribution of the four hot subdwarf He abundance classes and four groups in the $J_{z}-e$ diagram.
In Figure\,\ref{fig:fig10}, we plot the two regions defined by \cite{2003A&A...400..877P}.
Region A encompasses thin disk stars clustering in an area of low eccentricity and $J_{z}$ around $1800\,{\rm kpc\,km\,s^{-1}}$, while Region B encompasses thick disk stars having higher eccentricities and lower angular momenta.
Outside of these two regions, defined here as region C, halo star candidates are found.

As reported by \citet{2020ApJ...898...64L}, the majority of stars in the upper two panels of Figure\,\ref{fig:fig10} show a continuous distribution from Region A to Region B without any obvious dichotomy.
Only a few stars are found in Region C and He-rich stars with $\log{(y)}\ge-1$ have a very high fraction in this region.
The lower two panels of Figure\,\ref{fig:fig10} illustrate that Group $e$He-1 has a higher number fraction in Region C than Group $e$He-2, but the situation is the opposite in Region A.
Group $i$He-1 has a lower fraction than both Group $e$He-1 and $e$He-2 in Region B.
All stars in Group $i$He-2 have lower eccentricities and higher angular momenta and are found in Region A.

Table \ref{tab:tab2} displays the mean values and standard deviations of the orbital parameters: eccentricity, z-component of the angular momentum, normalised z-extent, maximum vertical amplitude, apocentre and pericentre.
Intermediate $i$He-rich stars with $-1\le\log(y)<0$ show the largest dispersion of the orbital parameters and $e$He-rich stars with $\log(y)\ge0$ display the second largest velocity dispersion.
The two groups of He-deficient sdB stars show similar velocity dispersions.
These results are in good agreement with the findings of \cite{2017MNRAS.467...68M} and support the different kinematic origins of the four He classes.
The trends in the dispersions of the orbital parameters of the four groups are consistent with the trends in the Galactic velocity distributions.

\subsection{Galactic Population classifications}\label{subsec:pop}
We adopted the classification scheme of \cite{2017MNRAS.467...68M} based on the $U-V$ diagram, $J_{z}-e$ diagram and the maximum vertical amplitude $z_{\rm max}$ to distinguish the Galactic populations of hot subdwarfs.
To ensure the correct population assignments, all orbits were visually inspected to supplement the automatic classifications.
As described by \cite{2020ApJ...898...64L}, stars within the $3\sigma$ thin disk contour in the $U-V$ diagram and Region A in the $J_{z}-e$ diagram are considered as thin disk stars.
Their orbits show only small excursions in Galactocentric distance $R$ and from the Galactic plane in the $Z$ direction, and they are confined in the $|z_{\rm max}|<1.5\,{\rm kpc}$ region.
Thick disk stars are situated within the $3\sigma$ thick disk contour and in Region B.
The extension of their orbits in $R$ and $Z$ direction are larger than that of thin disk stars, but do not cover the region of halo stars.
Halo stars lie outside Region A and B, as well as outside the $3\sigma$ thick disk contour.
Their orbits exhibit large differences in $R$ and $Z$.
Some halo stars have an extension in ($R$) larger than 18\,kpc, or the vertical distance from the Galactic plane ($Z$) larger than 6\,kpc.
In order to obtain the probabilities of the Galactic population memberships of each individual star, we performed Monte Carlo simulations for our sample using the parallaxes and proper motions provided by Gaia EDR3 and the RVs and their errors measured from LAMOST DR7 spectra.
Sets of 1000 Galactic space velocities and orbital parameters were produced for each individual star.
Then we obtained probabilities of Galactic population memberships by using the above classification scheme for each individual star and list them in Table \ref{tab:tab1}.

\begin{figure*}
\epsscale{1.2}
\plotone{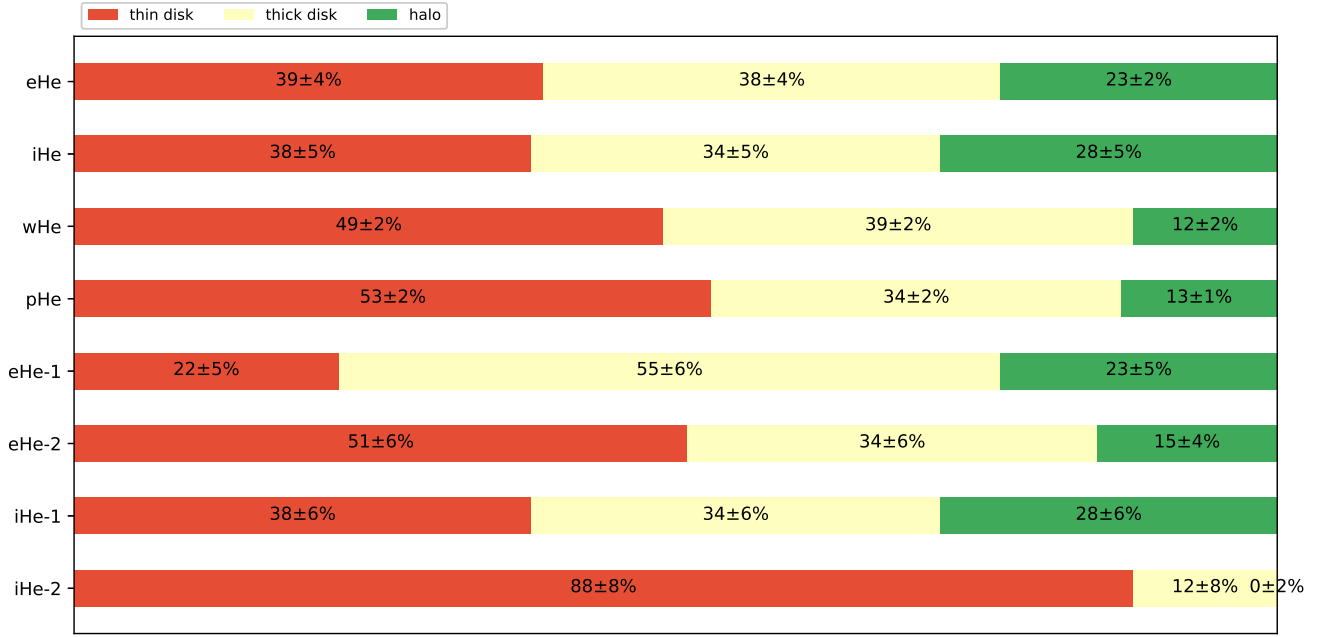}
\caption{The fractions of Galactic halo, thick disk and thin disk population corresponding to the four hot subdwarf helium abundance classes and four He-rich groups, respectively.
\label{fig:fig11}}
\end{figure*}

Table$\,$\ref{tab:tab4} gives the number of stars in the four hot subdwarf He classes and four He-rich groups classified as halo, thin and thick disk stars. Figure\,\ref{fig:fig11} illustrates the Galactic population (halo, thin and thick disk) fractions of the four He classes and four He-rich groups. The two He-deficient sdB groups have similar Galactic population fractions in the halo but exhibit $4\pm3\%$ differences in the thin and thick disk populations, while the two He-rich classes show about $12\pm3\%$ higher halo population fractions, and $13\pm3\%$ lower thin disk fractions.
Interestingly, the fraction of thick disk stars are between 34-40\% in all four He classes.
Comparisons of the He-rich and He-deficient classes show that the halo population fraction of the former is $12\pm3\%$ higher than that of the latter, and vice versa for thin disk population fractions.
This suggests a fundamental difference in the origin of He-rich and He-deficient stars.
Moreover, Group $e$He-1 shows $55\pm6\%$ thick disk population fraction, which is higher than those of thin disk and halo together, which indicates that the majority of stars in Group $e$He-1 belong to the thick disk.
Over half of the stars in Group $e$He-2 belong to the thin disk population and the halo population fraction is lower by $8\pm6\%$ than that of Group $e$He-1. The values of the thin and thick population fractions of Group $i$He-1 lie between Group $e$He-1 and Group $e$He-2. The value of the halo population fraction of Group $i$He-1 is significantly higher than that of Group $e$He-2.
Although Group $i$He-1 shows the largest value of the halo population fraction among the four He-rich groups, the difference between Groups $i$He-1 and $e$He-1 is not significant.
We found that Group $i$He-2 is the youngest population, that constitutes of $88\pm8\%$ thin disk and $12\pm8\%$ thick stars and no halo members.

\section{Discussions}\label{sec:dis}
\subsection{Comparisons with $\omega$ Cen }
\begin{figure*}
\epsscale{0.9}
\plotone{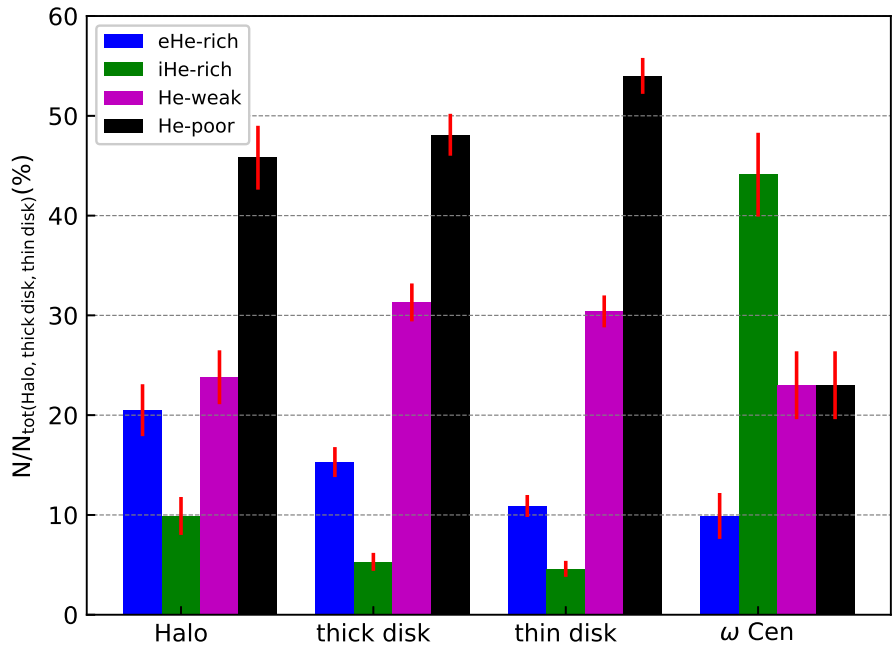}
\caption{
The fraction distribution comparisons of the four hot subdwarf helium abundance classes in the halo, thick disk, thin disk and the globular cluster $\omega$ Cen
\citep{2018A&A...618A..15L}.
\label{fig:fig12}}
\end{figure*}

\begin{figure*}
\epsscale{1.2}
\plotone{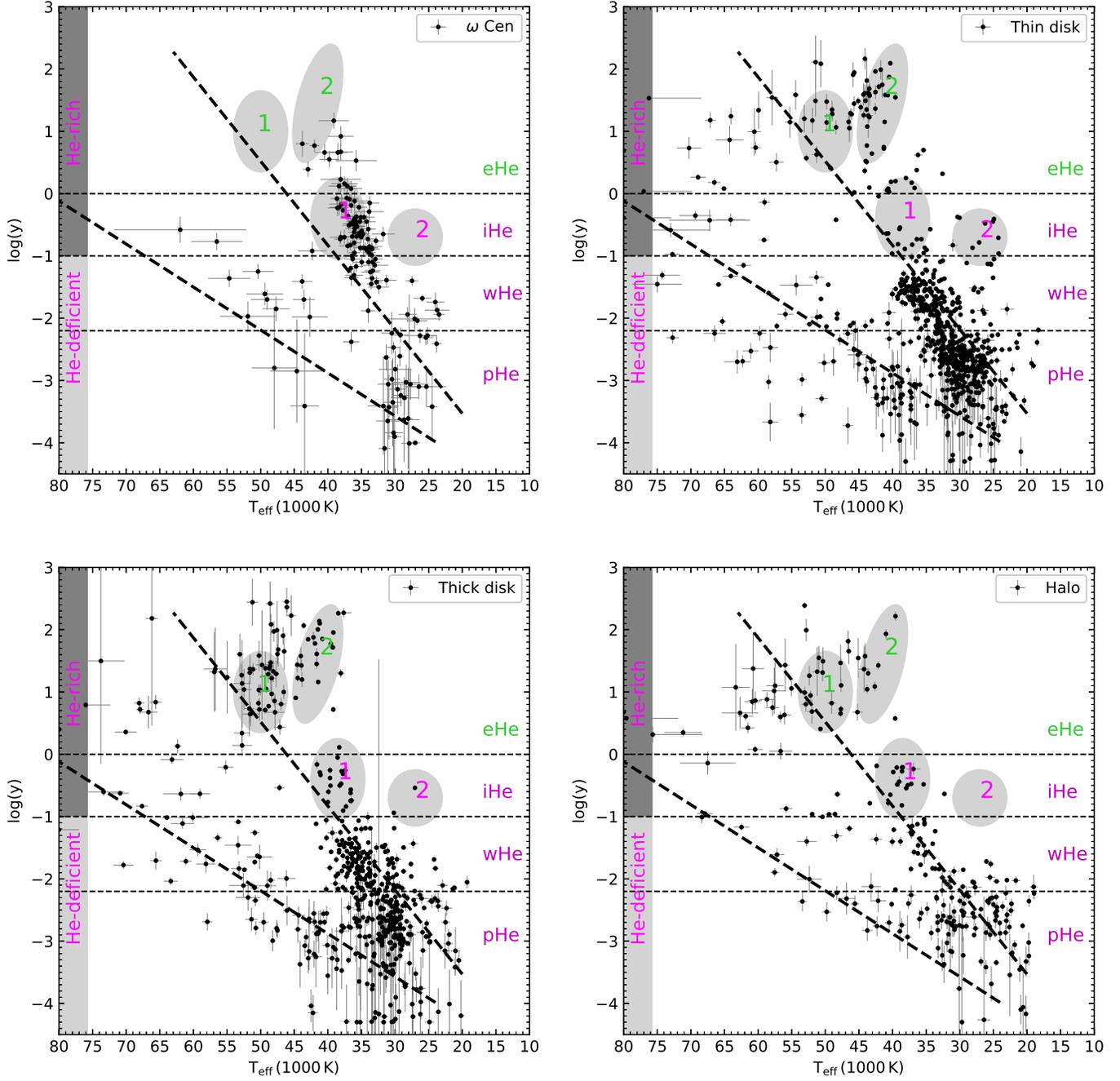}
\caption{$T_{\rm eff}-\log{(y)}$ diagrams for the four hot subdwarf helium abundance classes in the halo, thick disk, thin disk and the globular cluster $\omega$ Cen \citep{2018A&A...618A..15L}.
\label{fig:fig13}}
\end{figure*}

\cite{2018A&A...618A..15L} presented the spectroscopic properties of 152 hot subdwarfs in the Galactic globular cluster $\omega$ Cen, the largest sample of hot subdwarf stars in a globular cluster.
The sample is particularly interesting, because it allows for quantitative comparisons with hot subdwarf stars in the halo.
Figure\,\ref{fig:fig12} and Figure\,\ref{fig:fig13} illustrate the comparisons of hot subdwarf stars in the globular cluster $\omega$ Cen. Figure\,\ref{fig:fig12} exhibits the relative fractions of the four hot subdwarf helium groups in the thin disk, thick disk, halo and $\omega$ Cen.
In the halo, the relative fractions are $21\pm3\%$ for $e$He-rich stars with $\log{(y)}\ge0$, $10\pm2\%$ for $i$He-rich stars with $-1\le \log{(y)}<0$, $24\pm3\%$ He-weak stars with for $-2.2\le\log{(y)}<-1$, and $46\pm3\%$ for He-poor stars with $\log{(y)}<-2.2$, while in the globular cluster $\omega$ Cen the corresponding values are $10\pm2\%$, $44\pm4\%$, $23\pm3\%$ and $23\pm3\%$ for the four helium groups.
Only He-weak stars with $-2.2\le\log{(y)}<-1$ in the halo and $\omega$ Cen show approximately the same fractions, which indicates that the cluster environment has little influence on the formation of these stars.
The fractions of both $e$He-rich stars with $\log{(y)}\ge0$ and He-poor stars with $\log{(y)}<-2.2$ in the halo are two times larger than those in $\omega$ Cen, respectively.
However, the fraction of $i$He-rich stars with $-1\le \log{(y)}<0$ in the halo is only one-fourth of that in $\omega$ Cen.
These strongly suggest significant fractional differences between He-rich stars with $\log{(y)}\ge 0$ and $-1\le \log{(y)}<0$ and He-poor stars with $\log{(y)}<-2.2$ in the halo and in $\omega$ Cen.
This difference is attributed to the cluster's particular environment.
Figure\,\ref{fig:fig13} displays the distributions of hot subdwarfs in the $T_{\rm eff}-\log{(y)}$ diagram for the thin disk, thick disk and halo populations, as well as the distribution of hot subdwarf stars in $\omega$ Cen for comparison.
The Galactic field counterparts of $e$He-rich stars with $\log{(y)}\ge0$ in $\omega$ Cen appear in the thin disk, but they are not present in the thick disk and the halo.
Group $e$He-1 has no counterpart in $\omega$ Cen and there are only a few stars in $\omega$ Cen that are similar to the members of Group $e$He-2.

The overall trends of the relative fractions of the four hot subdwarf helium abundance classes are in agreement with our previous results \citep{2019ApJ...881....7L, 2020ApJ...898...64L}.
A study on the structure of the Milky Way \citep{2017ApJS..232....2X} demonstrated that the different Galactic populations (thin disk, thick disk and halo) represent different age stellar populations.
With the binary models of \cite{2002MNRAS.336..449H, 2003MNRAS.341..669H} the binary population synthesis calculations of \cite{2008A&A...484L..31H} gave the fractions of hot subdwarfs originated from three different formation channels (stable RLOF, CE ejection and the merger of double HeWDs) at various stellar population ages.
Although the exact values of the observed fractions cannot be matched with the predictions of binary population synthesis \citep{2003MNRAS.341..669H,2008A&A...484L..31H}, we find that the overall trends of the relative fractions are in good agreement with theory.

Past surveys of He-deficient sdB stars \citep{2003A&A...400..939E, 2005A&A...430..223L, 2007A&A...462..269S, 2009PhDT.......273H, 2012MNRAS.427.2180N, 2015A&A...577A..26G, 2016ApJ...818..202L, 2019ApJ...881....7L, 2018ApJ...868...70L, 2020ApJ...889..117L} showed that they can be divided into two groups by a gap in He abundance at $\log{(y)}=-2.2$. The formation channels of these two He-deficient groups are not fully understood in light of the currently available observations.
An analysis of long-period composite binary (sdB$+$F/G) candidates illustrated that they appear in both groups of He-deficient sdB stars, but show higher fractions among He-weak sdB stars with $-2.2 \le \log{(y)} < -1$ \citep{2012MNRAS.427.2180N}.
However, observations of larger samples \citep{2015MNRAS.450.3514K, 2015A&A...576A..44K} found that both short-period and long-period hot subdwarf binary systems occur in each sdB group.
In Figure\,\ref{fig:fig12}, our sample demonstrated that  the relative fractions of He-poor stars with $\log{(y)} < -2.2$ are in good agreement with the predictions of the CE ejection channel and those for He-weak sdB stars with $-2.2 \le \log{(y)} < -1$ agree with the predictions of the stable RLOF channel, if the excluded composite binary systems were all considered to have sdB stars with $-2.2 \le \log{(y)} < -1$.
Comparisons between hot subdwarf stars in the halo field and in the globular cluster $\omega$ Cen show that the cluster environment has no obvious influence on the fractions of He-weak stars with $-2.2 \le \log{(y)} < -1$, but the number of He-poor stars with $\log{(y)}<-2.2$ is only half in the cluster.
With the predictions of binary population synthesis of \cite{2008A&A...484L..31H} our sample indicates that He-poor stars with $\log{(y)}<-2.2$ are from the CE ejection channel, which is responsible for the production of short-period hot subdwarf binary systems.
He-weak stars with $-2.2 \le \log{(y)} < -1$ are from the stable RLOF channel, which creates long-period hot subdwarf binary systems.
Our sample show significant differences of these two He-deficient groups in the thin disk, thick disk and the halo.
Figure\,\ref{fig:fig12} exhibits that the He gap between the two He-deficient sdB groups clearly appears in the thin disk population, but disappears in the thick disk population, so that the two He-deficient sdB groups are no longer discernible in the thick disk and halo, which could be related to differences in the metallicity and age of the progenitor stars.
Further observations are needed to find the nature of hot subdwarfs in these two He-deficient sdB groups.

The relative fractions of $e$He-rich hot subdwarf stars with $\log{(y)}\ge0$ in the thin disk, thick disk and halo population are in good agreement with the predictions of the merger channel of double HeWDs.
They display two groups ($e$He-1 and $e$He-2) separated by a gap at effective temperature $T_{\rm eff}=46\,000$\,K in the $T_{\rm eff}-\log{g}$ and $T_{\rm eff}-\log{(y)}$ diagrams.
These two groups have been observed before and a systematic enhancement of carbon and nitrogen was found in Group $e$He-1 and 2, respectively \citep{2007A&A...462..269S, 2012MNRAS.427.2180N}.
The calculations of the merger of double HeWDs  \citep{2012MNRAS.419..452Z} demonstrated that the slow merger model produces nitrogen-rich hot subdwarf stars with a mass below $0.7M_{\odot}$ and the fast merger model produces carbon-rich stars with a mass over $0.7M_{\odot}$.
Therefore, we associate the hotter Group $e$He-1 with the fast merger model and the cooler Group $e$He-2 with the slow merger model.
Moreover, Figure\,\ref{fig:fig11} demonstrates that Group $e$He-1 has $22\pm5\%$ thin disk stars, $55\pm6\%$ thick disk stars and $23\pm5\%$ halo stars, while the corresponding fractions in Group $e$He-2 are $51\pm6\%$, $34\pm6\%$ and $15\pm4\%$, which indicates that Group $e$He-1 and $e$He-2 mainly formed in the thick disk and thin disk, respectively.
\cite{2012MNRAS.427.2180N} speculated that there is an evolutionary link from Group $e$He-2 to Group $e$He-1.
The relative number ratios between Group $e$He-1 and $e$He-2 are 0.42 in the thin disk, 1.6 in the thick disk and 1.5 in the halo, which implies that there must be other formation channels for Group $e$He-2 stars in the thin disk.
The comparisons between $e$He-rich hot subdwarf stars with $\log{(y)}\ge0$ in the halo field and in $\omega$ Cen show that Group $e$He-1 has no counterpart in $\omega$ Cen and Group $e$He-2 has only a few similar stars in $\omega$ Cen.
The fraction of $e$He-rich hot subdwarf stars with $\log{(y)}\ge0$ is more than $21\pm3\%$ in the halo field but less than $11\pm4\%$ in $\omega$ Cen, which strongly suggests that the particular cluster environment does not significantly contribute to the formation of $e$He-rich hot subdwarf stars with $\log{(y)}\ge0$.
Our comparisons of hot subdwarf stars in the thin disk, thick disk, halo, $\omega$ Cen, illustrate that the Galactic field counterparts of $e$He-rich stars with $\log{(y)}\ge0$ in $\omega$ Cen are the thin disk stars.
The fact that there are no counterparts of thick disk and halo stars in $\omega$ Cen, implies that there may exist a similar formation channel of hot subdwarfs with $\log{(y)}\ge0$ in the thin disk and in $\omega$ Cen.

The origin of $i$He-rich hot subdwarfs with $-1\le\log{(y)}<0$ is still a puzzle.
In Figure\,\ref{fig:fig12}, their relative fractions decrease from the halo to the thick disk, while they are nearly at equal frequency in the thick and thin disk.
Thanks to the improvements of Gaia EDR3 on parallaxes and proper motions, there are some differences with our previous results \citep{2019ApJ...881....7L,2020ApJ...898...64L} in the trend of the relative fraction distributions from the thick disk to the thin disk.
The overall tendency implies that $i$He-rich hot subdwarf stars with $-1\le\log{(y)}<0$ in the thin disk and the halo may have different formation channels and their relative contributions change with age.
The relative number ratios of $i$He-rich stars with $-1\le\log{(y)}<0$ with respect to $e$He-rich stars with $\log{(y)}\ge0$ are 0.50 in the halo, 0.36 in the thick disk and 0.40 in the thin disk, which are far larger than 0.2 predicted from main-sequence star (MS) and HeWD merger models \citep{2017ApJ...835..242Z}.
In the four He-groups, $i$He-rich subdwarf stars with $-1\le\log{(y)}<0$ show the largest dispersion of the Galactic space velocities and the highest halo population fraction.
These results suggest that the formation channels of $i$He-rich hot subdwarfs with $-1\le\log{(y)}<0$ are different from $e$He-rich subdwarf stars with $\log{(y)}\ge0$.
Moreover, $i$He-rich hot subdwarfs with $-1\le\log{(y)}<0$ display two groups ($i$He-1 and $i$He-2) in the $T_{\rm eff}-\log{g}$ and $T_{\rm eff}-\log{(y)}$ diagrams.
The majority of $i$He-rich stars with $-1\le\log{(y)}<0$ belongs to Group $i$He-1 and the minority belongs to Group $i$He-2.
The hot-flasher channels can explain these two groups, however, their kinematics show significant differences.
Group $i$He-2 has the smallest space velocity dispersion and consists of $88\pm8\%$ thin disk and $12\pm8\%$ thick disk stars.
The vast majority of stars in Group $i$He-2 belong to the thin disk.
Among the four He-rich groups, Group $i$He-1 has the highest halo population fraction.
These results indicate that the groups of $i$He-rich hot subdwarf stars with $-1\le\log{(y)}<0$ have different origins.
The comparison between $i$He-rich subdwarf stars with $-1\le\log{(y)}<0$ in the halo field and in $\omega$ Cen displays that the relative fractions in $\omega$ Cen are four times of that in the halo field, which suggests that the cluster environment can significantly increase the formation efficiency of $i$He-rich hot subdwarfs with $-1\le\log{(y)}<0$.

\subsection{Restricted sample for $D<1.5$ kpc}

\cite{2020ApJ...889..117L} pointed out and Figure \ref{fig:fig2} demonstrates that the LAMOST DR7 sample is representative of a volume-limited sample within 1.5 kpc.
The number of hot subdwarf stars in LAMOST DR7 and Gaia DR2 match within $D < 1.5$ kpc.
Beyond 1.5 kpc, more luminous stars become progressively overrepresented in our analysis.
Here we assume that all hot subdwarfs in the LAMOST footprint and within 1.5 kpc were observed, and we neglect all stars that are more distant. This allows restricting the fractional distributions to a volume-limited subsample of hot subdwarf stars.
We found that 590 of our stars are within 1.5 kpc, among which 467 are thin-disk stars, 115 are thick disk stars and only 8 are halo stars.
Figure 14 shows the fractional distributions of the Galactic populations of He classes within the restricted volume-limited sample.
The abundance classes in the disk populations show a remarkable similarity.
The $\omega$ Cen sample is markedly different, in particular, the He-poor and iHe-rich classes show the largest differences.
We found a low relative number of halo stars.
The error bars for the halo population clearly demonstrate that an investigation of the disk and halo populations cannot be made at a comparable precision from the currently available spectroscopic data.
For that, one must extend the sample to a larger volume and include more halo stars, however, for those fainter stars, obtaining accurate (spectroscopic or parallax-based) distances and high-quality spectroscopic measurements will be challenging.

Figure \ref{fig:fig15}  exhibits the relative fractional distributions of the 1587 Galactic halo, thick disk and thin disk population in Z distances for the four hot subdwarf helium abundance classes. The 1.5 kpc distance limit barely reaches the halo, therefore the relative errors for the halo population are large within the restricted volume-limited sample.
The comparison of the 1587 star sample to the volume-limited sample in Figure \ref{fig:fig16} reveals the extent of the disk populations.
The thin disk dominates the observed population over $|Z| = 1$ kpc and shows a nearly symmetric distribution to the mid-plane of the Galaxy.
The contribution of the thick disk population is as low as 10\% of all stars at the mid-plane, rises rapidly and over $|Z| = 1$ kpc it dominates over the thin disk.
The thick disk reaches its peak at $|Z| = 1.8\pm0.3$ kpc, where the thin disk is negligible.
The decline of the relative contribution of thick disk stars is much slower and their distribution resembles a Maxwell-Boltzmann distribution.
The halo population is very low, below 2\% in the mid-plane, and rises slowly.
Due to the low number statistics, the halo contribution can be determined much less accurately, but it dominates over the thick disk population beyond $|Z| = 3.5\pm0.5$ kpc.

The bright stars in a magnitude limited sample allow one to explore the extent of the disk populations on larger scales.
When the volume-limited sample is considered (bottom panels in Figure \ref{fig:fig16}, these bright stars are neglected and the double peak of the thick disk distribution remains poorly and partially covered.
The sample at $|Z| = 1.5$ kpc and within $D < 1.5$ kpc is limited to a few stars at very high galactic latitudes.
The halo contribution to the volume-limited sample is almost negligible (8 out of 590).
However, the relative fractions found in the volume-limited sample confirm the observations made in the 1587 star sample, 85\% of stars in the midplane belong to the thin disk and about 14\% belong to the thick disk and about 1\% belongs to the halo.
The two disk populations have equal contributions at $|Z| = 1$ kpc, just like in the 1587 star sample.
The volume-limited distribution looks boxier because the distance limit cuts off high $|Z|$ objects.
If we take the 467 thin disk stars and 115 thick disk stars in the volume-limited sample and observe in Figure \ref{fig:fig16} that the majority of these are within $|Z|<1$ kpc, we find the hot subdwarf number density is $39\pm6$ kpc$^{-3}$ in the thin disk and $9.6\pm1.5$ kpc$^{-3}$ in the thick disk, respectively.

We did not find significant differences among the correlations of atmospheric parameters of thin and thick disk stars.
This may be in part due to the atmospheric properties of hot subdwarfs being independent of their progenitor stars.
However, we expect a different bulk metallicity of these stars and, therefore, a difference in the total and core mass may exist and will be the subject of future work.

\section{Conclusions} \label{sec:conc}
We identified 1587 hot subdwarf stars with LAMOST DR7 observations from the Gaia DR2 hot subdwarf candidate catalog \citep{2019A&A...621A..38G}, among which 224 are confirmed as hot subdwarfs for the first time.
We obtained atmospheric parameters for these stars by fitting their LAMOST spectra with {\sc Tlusty/Synspec} non-LTE synthetic spectra.
Making use of the LAMOST radial velocities and the Gaia EDR3 parallaxes and proper motions, we presented the Galactic space positions, velocity vectors and orbital parameters.
By combining the Galactic velocity components with orbital parameters, we calculated the Galactic population memberships of the stars and made a comparison with the Galactic globular cluster $\omega$ Cen.
Following our previous works \citep{2019ApJ...881....7L, 2020ApJ...898...64L}, these stars were divided into four classes depending on their He abundances. We summarize our results in Fig\,\ref{fig:fig17} as follows:

\begin{enumerate}
  \item
  The atmospheric parameters of our sample reproduce and confirm the two known He sequences of sdB stars in the $T_{\rm eff}-\log{(y)}$ diagram.
  According to our He classification scheme, our sample is divided into four He hot subdwarf classes in the $T_{\rm eff}-\log{g}$ and $T_{\rm eff}-\log{(y)}$ diagrams.
  Two groups ($e$He-1 and $e$He-2) can be outlined for $e$He-rich stars with $\log{(y)}\ge0$ and another two groups ($i$He-1 and $i$He-2) for $i$He-rich stars with $-1\le\log{(y)}<0$.
  Group $i$He-2 is a newly discovered group.
  Our comparisons with theoretical hot subdwarf formation models illustrate that the hot-flasher evolutionary tracks \citep{2008A&A...491..253M} can match Groups $e$He-2, $i$He-1 and $i$He-2, and the double WD merger model \citep{2012MNRAS.419..452Z} can explain the properties of Group $e$He-1 and $e$He-2.
  We associate Group $e$He-1 with the fast merger channel and Group $e$He-2 with the slow merger channel.
  \item
  The four He classes exhibit some noticeable differences in their space distributions, the Galactic velocity distributions and the Galactic orbital parameter distributions, which are clearly reflected in their Galactic population classifications.
  The two He-deficient classes show very similar contributions to the halo populations, while the two He-rich classes show no differences in their relative fractions in the thick disk and the halo.
  The halo population fractions of both He-rich classes are higher by $12\pm3\%$ than those of the He-deficient classes.
  The situation is the opposite for the thin disk population fractions, which suggests that He-rich and He-deficient stars have different formation channels.
  The two groups ($e$He-1 and $e$He-2) of He-rich stars with $\log{(y)}\ge0$ show significant differences in all three population fractions.
  Group $e$He-1 constitutes of $22\pm5\%$ thin disk stars, $55\pm6\%$ thick disk stars and $23\pm5\%$ halo stars, while the corresponding fractions in Group $e$He-2 are $51\pm6\%$ in the thin disk, $34\pm6\%$ in the thick disk and $15\pm4\%$ in the halo.
  These suggest that over half of the stars in Group $e$He-1 are halo stars, but over half of stars in Group $e$He-2 are thin disk stars.
  It is possible that these two groups are from different formation channels. The values of the thin and thick disk population fractions of Group $i$He-1 are between those of Group $e$He-1 and $e$He-2. The halo population of Group $i$He-1 is significantly higher than that of $e$He-2 but it has no significant difference between Groups $i$He-1 and $e$He-2.
  Group $i$He-1 is completely different from Group $i$He-2 in the Galactic population fractions.
  The latter has no halo stars and shows $88\pm8\%$ thin disk population fraction and $12\pm8\%$ thick disk population fraction, which implies that Groups $i$He-1 and $i$He-2 have different formation channels.
  \item
  The relative fractions of the four hot subdwarf helium classes in the halo, thin disk and thick disk can be largely matched with our previous results \citep{2019ApJ...881....7L,2020ApJ...898...64L}, which appears to support the predictions of binary population synthesis \citep{2003MNRAS.341..669H,2008A&A...484L..31H}.
  He-weak stars with $-2.2\le\log{(y)}<-1$ are likely originate from the stable RLOF channel, He-poor stars with $\log{(y)}<-2.2$ are from the CE ejection channel and $e$He-rich stars with $\log{(y)}\ge0$ are from the merger channel of double HeWDs.
  The comparisons between hot subdwarf stars in the halo and the globular cluster $\omega$ Cen show that of the four He classes, except for the He-weak class with $-2.2 \le \log{(y)}<-1$, have a significant difference in their relative population fractions.
  Both the $e$He-rich class with $\log{(y)}\ge0$ and the He-poor class with $\log{(y)}<-2.2$ are two times smaller in the halo field than in $\omega$ Cen.
  The relative fraction of $i$He-rich stars with $-1 \le \log{(y)}<0$ in $\omega$ Cen is four times of that in the halo field.
  These results suggest that the particular cluster environment has little effect on the formation of He-deficient stars formed by the RLOF channel, but it does significantly cut the formation of hot subdwarf stars through the double WD merger and CE ejection channels and significantly contribute to the formation of $i$He-rich stars with $-1 \le \log{(y)}<0$ in spite of their puzzling origin.
  We find that $e$He-rich stars with $\log{(y)}\ge0$ discovered in $\omega$ Cen are missing in the thick disk and halo, but they appear in the thin disk, which indicates that these stars may have a similar formation channel in the thin disk and $\omega$ Cen.

  \item
  We have found substantial differences among hot subdwarf He abundance classes and groups.
  By combining these differences with kinematic properties, we see a fragmentation of the galactic hot subdwarf population.
  There are different dichotomies between the thick and thin disk populations, and the halo and $\omega$ Cen populations of stars.
  It is the scope of future research to find if the observed dichotomies are present due to differences in stellar and binary evolution in different galactic environments, or due to a dilution of disk stars by a past major merger of the Milky Way with a satellite galaxy.
  By simple population membership considerations, we found a hot subdwarf number density of $39\pm6 $kpc$^{-3}$ in the thin disk and $9.6\pm1.5$ kpc$^{-3}$ in the thick disk for the solar neighborhood, respectively.
  We did not find significant differences among the correlations of atmospheric parameters of thin and thick disk stars.
  However, we expect a different bulk metallicity of these stars and a difference in the total and core mass may exist.

\end{enumerate}

\acknowledgments
The research presented here is supported by
the National Natural Science Foundation of China under
grant no. U1731111, 12003022, 12090040, 12090043 and 11521303  and the Sichuan Science and Technology Program no.2020YFSY0034.
P.N. acknowledges support from the Grant Agency of the Czech Republic (GA\v{C}R 18-20083S).
Guoshoujing Telescope (the Large Sky Area Multi-Object Fiber Spectroscopic
Telescope LAMOST) is a National Major Scientific
Project built by the Chinese Academy of Sciences.
Funding for the project has been provided by the National
Development and Reform Commission. LAMOST
is operated and managed by the National Astronomical
Observatories, Chinese Academy of Sciences.
This work has made use of data from the European Space Agency (ESA) mission
{\it Gaia} (\url{https://www.cosmos.esa.int/gaia}), processed by the {\it Gaia}
Data Processing and Analysis Consortium (DPAC,
\url{https://www.cosmos.esa.int/web/gaia/dpac/consortium}). Funding for the DPAC
has been provided by national institutions, in particular the institutions
participating in the {\it Gaia} Multilateral Agreement.
This research has used the services of \mbox{\url{www.Astroserver.org}}.
\software{\,astropy \citep{2013A&A...558A..33A, 2018AJ....156..123A}, \,TOPCAT (v4.6; \citealt{2005ASPC..347...29T,2019ASPC..523...43T}), \,galpy \citep{2015ApJS..216...29B}}.

\begin{figure*}
\epsscale{0.9}
\plotone{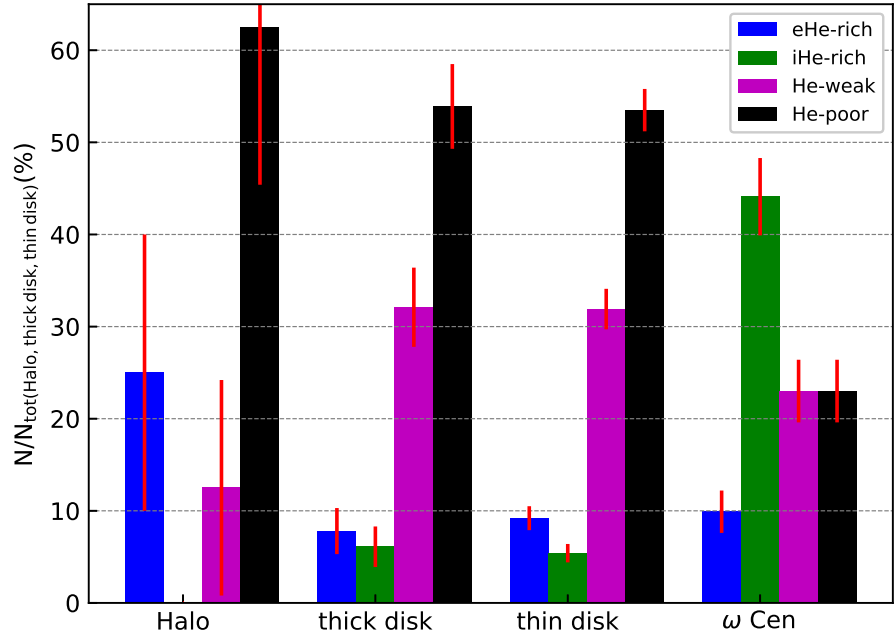}
\caption{
The fraction distribution comparisons of the four hot subdwarf helium abundance classes in the halo, thick disk, thin disk for the restricted sample for $D<1.5$ kpc and the globular cluster $\omega$ Cen
\citep{2018A&A...618A..15L}.}
\label{fig:fig14}
\end{figure*}

\begin{figure*}
\epsscale{1.2}
\plotone{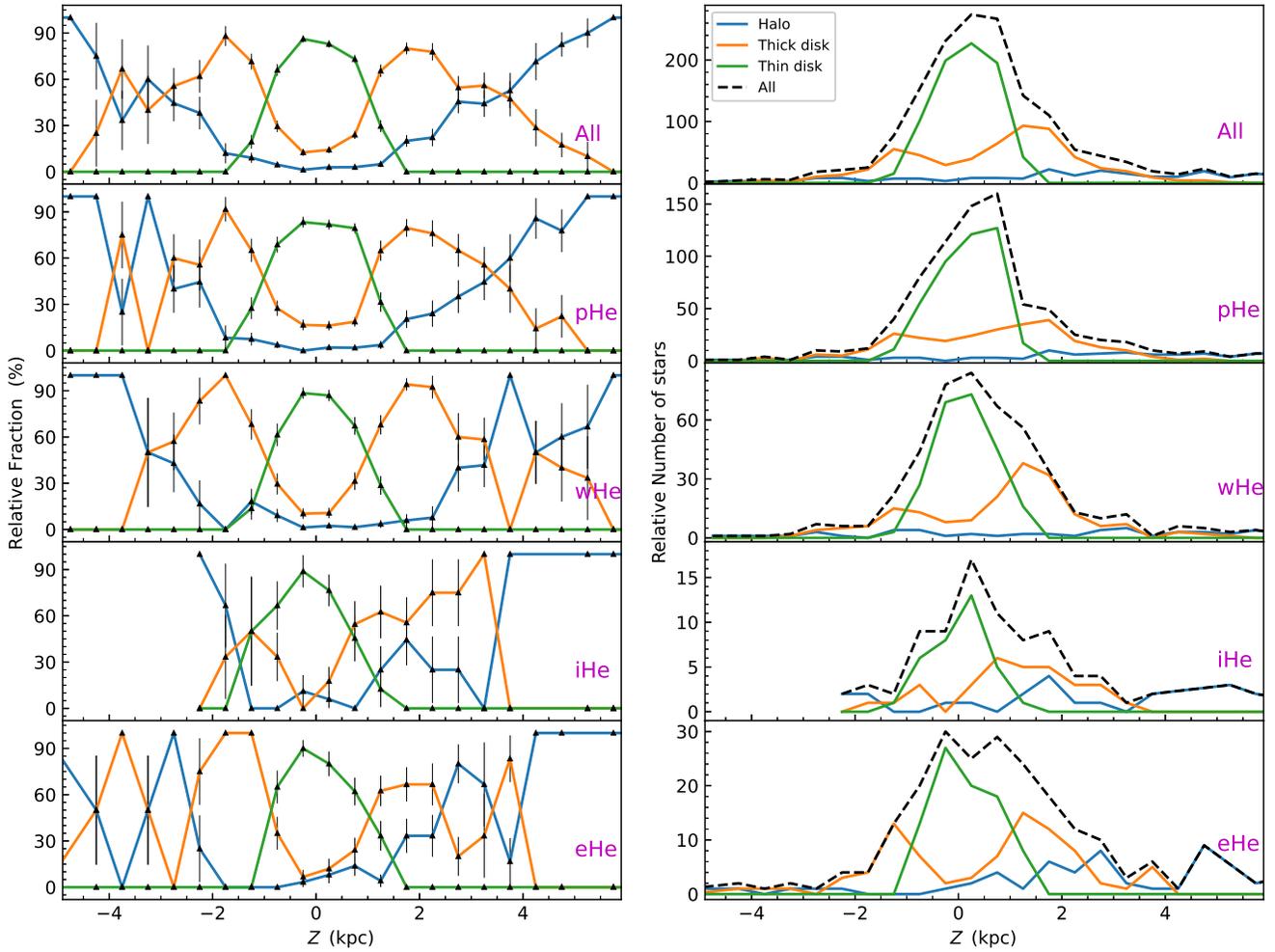}
\caption{Relative contributions of the four hot subdwarf helium abundance classes to the Galactic halo, thick and thin disk populations as a function of $Z$ distance.}
\label{fig:fig15}
\end{figure*}

\begin{figure*}
\epsscale{1.2}
\plotone{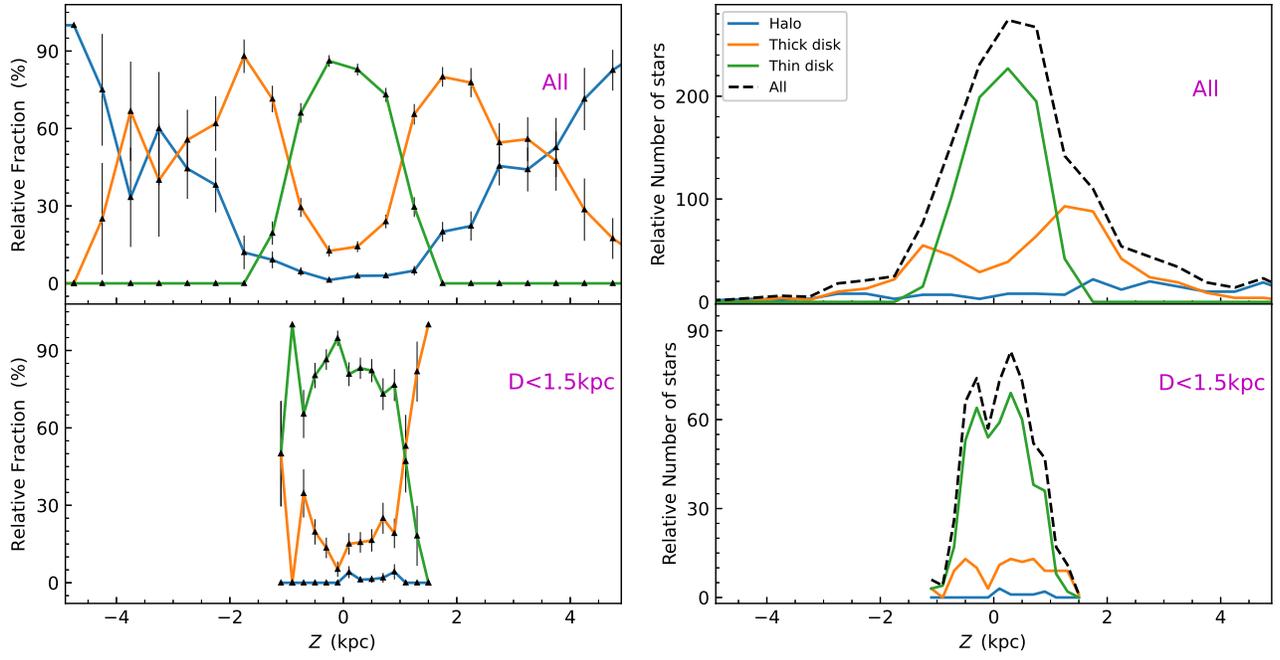}
\caption{The comparisons of the total sample to the restricted sample for $D<1.5$ kpc for the relative contributions of the hot subdwarf stars to the Galactic halo, thick and thin disk populations as a function of $Z$ distance.}
\label{fig:fig16}
\end{figure*}

\begin{figure*}
\epsscale{0.9}
\plotone{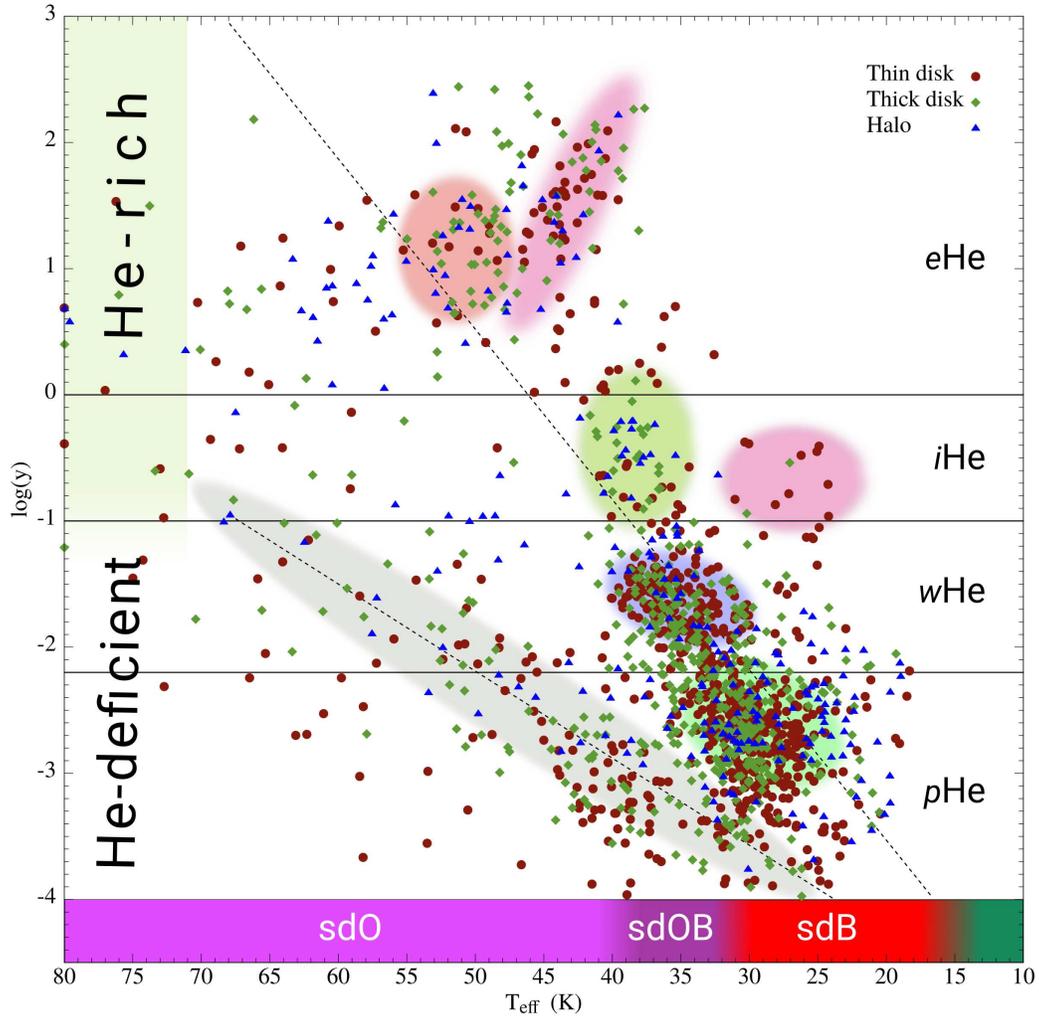}
\caption{
Atmospheric parameters and galactic population memberships of 1587 hot subdwarf stars in the He abundance classification scheme outlined in Fig\,\ref{fig:fig4}.
\label{fig:fig17}}
\end{figure*}

\begin{deluxetable*}{rlll}
\tablecaption{Atmospheric parameters, space positions, orbital parameters and galactic velocities for 1587 hot subdwarf stars identified in Gaia DR2 with spectra from LAMOST DR7. \label{tab:tab1}}
\tablewidth{0pt}
\tabletypesize{\scriptsize}
\tablehead{
\colhead{Num} & \colhead{Label}    & \colhead{Explanations}
}
\startdata
1  &   LAMOST          & LAMOST target\\
2  &      $RAdeg$      & Barycentric Right Ascension (J2000) \tablenotemark{(1)}\\
3  &      $DEdeg$      & Barycentric Declination (J2000) \tablenotemark{(1)}\\
4  &    $T_{\rm eff}$  & Stellar effective temperature\\
5  & $e\_T_{\rm eff}$  & Standard error in $T_{\rm eff}$ \\
6  &         $\log g$  & Stellar surface gravity\\
7  &      $e\_\log g$  & Standard error of Stellar surface gravity \\
8  &        $\log(y)$  & Stellar surface He abundance $y=n{\rm He}/n{\rm H}$\\
9  &     $e\_\log(y)$  & Standard error in $\log(y)$ \\
10 &           $pmRA$  & Gaia EDR3 proper motion in RA \\
11 &        $e\_pmRA$  & Standard error $pmRA$ \\
12 &           $pmDE$  & Gaia EDR3 proper motion in DE \\
13 &        $e\_pmDE$  & Standard error in $pmDE$\\
14 &              $D$  & Gaia EDR3 stellar distance\\
15 &           $e\_D$  & Standard error in stellar distance\\
16 &           $RVel$  & Radial velocity from LAMOST spectra\\
17 &        $e\_RVel$  & Standard error in radial velocity\\
18 &              $X$  & Galactic position towards Galactic center\\
19 &           $e\_X$  & Standard error in $X$\\
20 &              $Y$  & Galactic position along Galactic rotation\\
21 &           $e\_Y$  & Standard error of $Y$\\
22 &              $Z$  & Galactic position towards north Galactic pole\\
23 &           $e\_Z$  & Standard error of $Z$\\
24 &              $U$  & Galactic radial velocity positive towards Galactic center\\
25 &           $e\_U$  & Standard error in $U$ \\
26 &              $V$  & Galactic rotational velocity along Galactic rotation \\
27 &           $e\_V$  & Standard error in $V$\\
28 &              $W$  & Galactic velocity towards north Galactic pole\\
29 &           $e\_W$  & Standard error in $W$ \\
30 &         $R_{ap}$  & Apocenter radius \tablenotemark{(2)} \\
31 &      $e\_R_{ap}$  & Standard error in $R_{ap}$\\
32 &       $R_{peri}$  & Pericenter radius \tablenotemark{(2)}\\
33 &    $e\_R_{peri}$  & Standard error in $R_{peri}$\\
34 &    $z_{\rm max}$  & Maximum vertical height \tablenotemark{(2)}\\
35 & $e\_z_{\rm max}$  & Standard error in $z_{\rm max}$\\
36 &              $e$  & Eccentricity \tablenotemark{(2)}\\
37 &           $e\_e$  & Standard error in $e$\\
38 &      $J_{\rm z}$  & Z$-$component of angular momentum \tablenotemark{(2)}\\
39 &   $e\_J_{\rm z}$  & Standard error in $J_{\rm z}$ \\
40 &      $z_{\rm n}$  & Normalised z-extent of the orbit \tablenotemark{(2)}\\
41 &   $e\_z_{\rm n}$  & Standard error in $z_{\rm n}$\\
42 &   Groupid         & Group name for He-rich stars \tablenotemark{(3)} \\
43 &     Pops          & Population classification \tablenotemark{(4)}\\
44 & $P_{\rm TH}$      & probability in thin disk\\
45 & $P_{\rm TK}$      & probability in thick disk\\
46 & $P_{\rm H}$       & probability in halo\\
47 & Newsample         & New hot subdwarf star identified in LAMOST DR7 \tablenotemark{(5)}\\
\enddata
\tablenotetext{(1)}{At Epoch 2000.0 (ICRS).}
\tablenotetext{(2)}{Form the numerical orbit integration.}
\tablenotetext{(3)}{1$=$$e$He-1; 1$=$$e$He-2; 3$=$$i$He-1; 4$=$$i$He-2.}
\tablenotetext{(4)}{H$=$Halo; TK$=$thick disk; TH$=$thin disk.}
\tablenotetext{(5)}{Y$=$Yes; N$=$No.}
\tablecomments{The full table can be found in the online version of the paper.}
\end{deluxetable*}

\begin{deluxetable*}{rlll}
\tablecaption{The excluded targets from the Gaia DR2 catalogue of hot subdwarf star candidates with spectra from LAMOST DR7. \label{tab:tab2}}
\tablewidth{0pt}
\tabletypesize{\scriptsize}
\tablehead{
\colhead{Num} & \colhead{Label}    & \colhead{Explanations}
}
\startdata
1  &   LAMOST          & LAMOST target\\
2  &   GaiaEDR3        & Unique Gaia EDR3 source identifier \\
3  &      $RAdeg$      & Barycentric Right Ascension (J2000) \tablenotemark{(1)}\\
4  &      $DEdeg$      & Barycentric Declination (J2000) \tablenotemark{(1)}\\
5  &        $Plx$      & Gaia EDR3 stellar parallax\\
6  &     $e\_Plx$      & Standard error in $Plx$ \\
7  &       $pmRA$      & Gaia EDR3 proper motion in RA \\
8  &    $e\_pmRA$      & Standard error $pmRA$ \\
9  &       $pmDE$      & Gaia EDR3 proper motion in DE \\
10 &    $e\_pmDE$      & Standard error in $pmDE$\\
11 &       $Gmag$      & Gaia EDR3 G$-$band magnitude\\
12 &      $BP-RP$      & Gaia EDR3 $BP-RP$ colour \\
13 &      SpClass      & Spectra classification (photBpMeanMag$-$photRpMeanMag) \tablenotemark{(2)}\\
\enddata
\tablenotetext{(1)}{At Epoch 2000.0 (ICRS).}
\tablenotetext{(2)}{spectral classes are as follows:
                  BHB $=$ Blue Horizontal Branch star;
                  CV $=$ Cataclysmic Variable star;
                  DA $=$ white dwarfs with H lines only;
                  DB $=$ white dwarfs with neutral He lines only;
                  MS$-$B $=$ B type main sequence star;
                  MS$-$A $=$ A type main sequence star;
                  sdA $=$ A type subdwarf star;
                  sd$+$MS $=$ hot subdwarf$-$Main Sequence binary candidates;
                  UNK $=$ Bad spectra which we were unable to classify;
                  Unreliable $=$ Spectra with low signal.}
\tablecomments{The full table can be found in the online version of the paper.}
\end{deluxetable*}

\begin{longrotatetable}
\begin{deluxetable*}{lclclclclclclclclclclclc}
\tablecaption{Mean values and dispersion (standard deviation) of the Galactic velocities $WVUV_{\rm tot}$ ($V_{\rm tot}=\sqrt{U^{2}+V^{2}+W^{2}}$) and the Galactic orbital parameters: eccentricity ($e$),  z-component of the angular momentum $J_{\rm z}$, normalised z-extent ($z_{\rm n}$),  maximum vertical amplitude ($z_{\rm max}$), apocentre ($R_{\rm ap}$) and pericentre  ($R_{\rm peri}$) for the four hot subdwarf helium abundance classes and four He-rich groups. \label{tab:tab3}}
\tablewidth{1500pt}
\tablehead{
\colhead{Subsample} & \colhead{N} &
\colhead{$\bar{U}$} & \colhead{$\sigma_{U}$} &
\colhead{$\bar{V}$} & \colhead{$\sigma_{V}$} &
\colhead{$\bar{W}$} & \colhead{$\sigma_{W}$} &
\colhead{$\overline{V_{\rm tot}}$} & \colhead{$\sigma_{V_{\rm tot}}$} &
\colhead{$\overline{e}$}           & \colhead{$\sigma_{e}$}&
\colhead{$\overline{J_{\rm z}}$}           & \colhead{$\sigma_{J_{\rm z}}$}&
\colhead{$\overline{z_{\rm n}}$}   & \colhead{$\sigma_{z_{\rm n}}$}&
\colhead{$\overline{z_{\rm max}}$} & \colhead{$\sigma_{z_{\rm max}}$}&
\colhead{$\overline{R_{\rm ap}}$}      & \colhead{$\sigma_{R_{\rm ap}}$}&
\colhead{$\overline{R_{\rm peri}}$}    & \colhead{$\sigma_{R_{\rm peri}}$}
}
\startdata
All stars  &1587 & 14 & 62 & 198 & 47 & -1  & 38 & 214 & 42 & 0.24 & 0.16 & 1763.74 & 598.04 & 0.13 & 0.10 & 1.30 & 0.94 & 10.21 & 1.91 & 6.12 & 2.61\\
$p$He        & 803 & 15 & 61 & 203 & 44 & -1  & 35 & 216 & 40 & 0.23 & 0.14 & 1824.73 & 525.26 & 0.12 & 0.09 & 1.21 & 0.84 & 10.18 & 1.90 & 6.25 & 2.47\\
$w$He        & 472 & 12 & 58 & 198 & 47 & -1  & 38 & 212 & 41 & 0.24 & 0.15 & 1783.13 & 589.84 & 0.13 & 0.10 & 1.33 & 0.97 & 10.13 & 1.74 & 6.23 & 2.58\\
$i$He        & 90  & 8  & 82 & 173 & 72 & -10 & 48 & 206 & 52 & 0.36 & 0.26 & 1456.57 & 826.99 & 0.16 & 0.14 & 2.20 & 2.09 & 10.02 & 1.92 & 5.35 & 2.88\\
$e$He        &222  & 15 & 87 & 181 & 58 & -2  & 40 & 209 & 45 & 0.34 & 0.26 & 1550.70 & 866.36 & 0.18 & 0.14 & 1.72 & 1.32 & 10.45 & 2.11 & 5.74 & 3.01\\
$e$He-1      & 69  & -8 & 90 & 166 & 70 & -15 & 49 & 204 & 51 & 0.37 & 0.28 & 1429.64 & 938.69 & 0.23 & 0.16 & 2.34 & 1.56 & 10.58 & 2.36 & 5.36 & 3.15\\
$e$He-2      & 71  & 13 & 76 & 196 & 42 & -1  & 27 & 214 & 35 & 0.25 & 0.18 & 1769.81 & 494.10 & 0.11 & 0.08 & 1.11 & 0.73 & 10.03 & 1.72 & 6.05 & 2.51\\
$i$He-1      & 65  & 7  & 82 & 167 & 70 & -9  & 55 & 203 & 50 & 0.38 & 0.28 & 1358.52 & 816.37 & 0.19 & 0.17 & 2.40 & 2.17 & 9.64  & 1.47 & 5.07 & 2.84\\
$i$He-2      & 17  & 12 & 50 & 238 & 27 & -3  & 17 & 244 & 24 & 0.14 & 0.06 & 2244.47 & 350.01 & 0.06 & 0.03 & 0.68 & 0.43 & 11.09 & 1.96 & 8.32 & 1.57\\
\enddata
\end{deluxetable*}
\end{longrotatetable}

\begin{deluxetable*}{lclclclclc}
\tablecaption{Population classifications for the four hot subdwarf helium abundance classes and four He-rich groups.\label{tab:tab4}}
\tablewidth{0pt}
\tablehead{
\colhead{Subsample}  &
\colhead{N}          &
\colhead{Thin Disk}  &
\colhead{Thick Disk} &
\colhead{Halo}
}
\startdata
All stars        & 1587  &  779   &  568   & 240  \\
$p$He              & 803   &  426   &  269   & 108   \\
$w$He              & 472   &  233   &  183   & 56   \\
$i$He              & 90    &  34    &  31    & 25   \\
$e$He              & 222   &  86    &  85    & 51   \\
$e$He-1            & 69    &  15    &  38    & 16   \\
$e$He-2            & 71    &  36    &  24    & 11   \\
$i$He-1            & 65    &  25    &  22    & 18   \\
$i$He-2            & 17    &  15    &  2     & 0    \\
\enddata
\end{deluxetable*}

\end{document}